\documentclass[manuscript,screen,nonacm]{acmart}

\usepackage{float}
\usepackage{stfloats}
\usepackage{tabularx}
\usepackage{array}
\usepackage{threeparttable}
\usepackage[utf8]{inputenc}
\usepackage{textgreek}
\usepackage{longtable}
\usepackage{soul, color, xcolor}
\usepackage{xcolor}
\usepackage{framed}

\definecolor{highlightyellow}{RGB}{255,255,153}

\AtBeginDocument{%
  }

\setcopyright{none}
\copyrightyear{2026}
\acmYear{2026}

\begin{document}

\title{WhiteTesseract: Reframing the Interpretation of Cultural Heritage through XR and Conversational AI}

\author{Jingjing Li}
\email{li@digitalnature.slis.tsukuba.ac.jp}
\orcid{0000-0002-6524-3105}
\affiliation{
  \institution{University of Tsukuba}
  \city{Tsukuba}
  \country{Japan}
}
\author{Zhi Liu}
\orcid{0000-0002-8706-6269}
\affiliation{
  \institution{University of Tsukuba}
  \city{Tsukuba}
  \country{Japan}
}
\author{Xiyao Jin}%
\orcid{0009-0009-5638-7973}
\affiliation{
  \institution{University of Tsukuba}
  \city{Tsukuba}
  \country{Japan}
}
\author{Tatsuki Fushimi}
\orcid{0000-0003-3944-0014}
\affiliation{
  \institution{University of Tsukuba}
  \city{Tsukuba}
  \country{Japan}
}
\author{Yoichi Ochiai}
\orcid{0000-0002-4690-5724}
\affiliation{
  \institution{University of Tsukuba}
  \city{Tsukuba}
  \country{Japan}
}

\renewcommand{\shortauthors}{Li et al.}

\begin{teaserfigure}
  \centering
  \includegraphics[width=\linewidth]{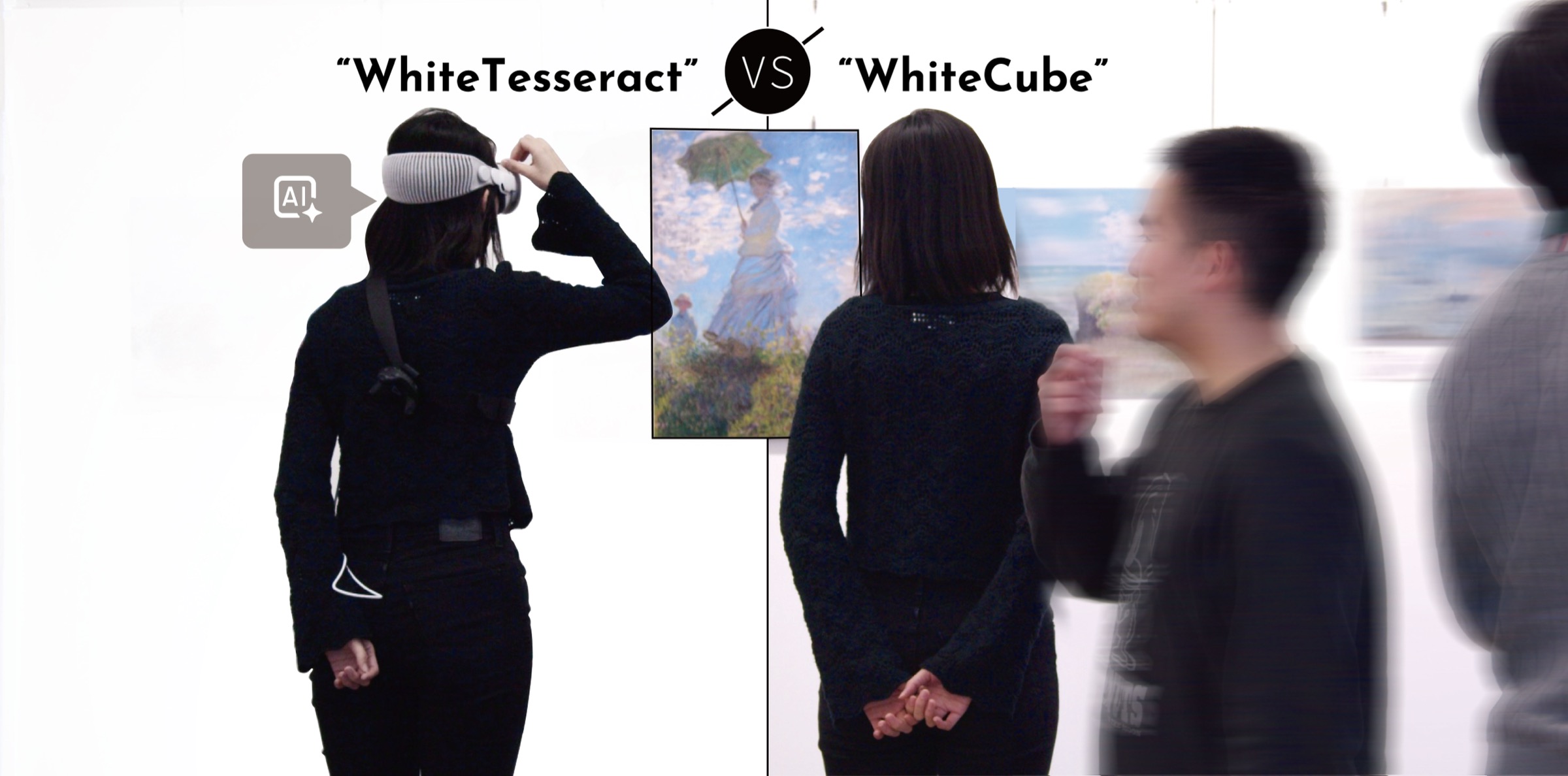}
  \caption{WhiteCube vs WhiteTesseract. Comparison between the traditional gallery condition (WhiteCube, right) and the XR-AI condition (WhiteTesseract, left).}
  \Description{A woman and a girl in white dresses sit in an open car.}
\end{teaserfigure}

\begin{abstract}

Cultural heritage exhibitions often struggle to sustain attention and support reflective engagement. While physical exhibitions incorporate various interpretive aids, such as labels, audio guides, and tactile features, they often deliver fixed content that lacks adaptability to individual backgrounds or curiosity. Their effectiveness is highly dependent on a visitor’s Personal Context, prior knowledge, cultural literacy, and interpretive confidence. Meanwhile, digital exhibitions often prioritize convenience and accessibility but risk weakening the Physical and Social Contexts that define embodied cultural experience.

WhiteTesseract addresses this gap by enabling in-situ interpretation through high-resolution XR and conversational AI. The system integrates spatial intelligence via artwork recognition and perceptual modulation to allow visitors to selectively reduce environmental distractions (via diminished reality) and engage in context-aware dialogue (via large language models). The goal is to preserve the richness of the physical and social environment while providing a flexible space for personal reflection, enhancing Personal Context without compromising physical authenticity.

We deployed the system in a Claude Monet exhibition and conducted a controlled user study with 27 recruited participants, among which 26 completed the entire experiment and were included in the analysis. Quantitative results showed that WhiteTesseract modulation significantly increased average viewing duration from 35.3 to 98.3 seconds (\textit{p} < 0.001). In parallel, analysis of 529 visitor–AI interactions revealed that 60\% extended beyond factual queries to include analytical, emotional, and comparative inquiries. These findings demonstrate how XR and AI can enrich the physical exhibition experience by supporting deeper, more personalized engagement without displacing the embodied value of cultural heritage. We discuss technical and social constraints for real-world deployment and limitations of our controlled experimental setting.

\end{abstract}




\received{20 February 2007}
\received[revised]{12 March 2009}
\received[accepted]{5 June 2009}

\maketitle

\section{INTRODUCTION}

Over the past decade, digital technologies have significantly reshaped how cultural institutions present and interpret cultural heritage materials. While virtual exhibitions and immersive environments have enabled access beyond physical constraints, studies continue to emphasize the unique perceptual, spatial, and social affordances of physical exhibition spaces in shaping visitor experiences \cite{maxwell2002museums, forrest2013museum}. Among these, the “White Cube” gallery, a neutral, distraction-minimizing spatial model, has remained a dominant format for encouraging focused and contemplative engagement with art \cite{o1999inside}, especially in the presentation of visual art as cultural heritage. However, its strengths in visual isolation can also limit multimodal interaction and individualized interpretive depth \cite{brieber2015white}.

With the emergence of \textbf{high-resolution extended reality (XR)} devices \cite{chuah2018and, alnagrat2022review, bhaskaran2024immersive}, a new question arises: \textbf{how might such technologies augment, rather than substitute, the real-world cultural exhibition experience?} While physical exhibitions offer authenticity and embodied presence, they also present limitations. Visitors must navigate visually dense environments, time constraints, and limited contextual information \cite{brieber2014art, bitgood2009museum}, factors that can shorten viewing durations and constrain interpretation to surface-level observations. In minimalist gallery formats such as the White Cube, the reduction of visual noise supports focused looking yet often lacks mechanisms to adapt to individual curiosity or facilitate a deeper understanding of the artwork’s context, meaning, or technique \cite{bishop2005installation}.

These gaps suggest two dimensions where technological augmentation may be useful: attentional focus, referring to a visitor’s ability to control what is perceived within a complex sensory field; and cognitive engagement, referring to the depth of thought, questioning, and interpretation that emerges during the viewing process. While most exhibition environments offer limited support for either XR or AI technologies, they can be designed to intervene at both perceptual and interpretive levels. In our system, the attentional focus is supported through diminished reality (DR) techniques that allow users to reduce visual distractions by selectively masking surrounding elements and auditory distractions through spatial noise cancellation \cite{mori2017survey, cheng2022towards}. These combined interventions aim to create a more isolated perceptual space that encourages sustained attention and concentration. Rather than adding visual or auditory load, the system acts to simplify the sensory environment, helping visitors dwell longer and think more deeply, without disrupting the aesthetic or curatorial integrity of the exhibition.

In this work, we present WhiteTesseract, a system that builds upon the White Cube paradigm by introducing programmable spatial and cognitive extensions via XR and AI technologies. The system integrates two key features:
(1) a diminished reality interface that allows visitors to dynamically reduce both visual and auditory distractions around individual artworks, creating what we term an adjustable attention space and
(2) a language model-based dialogue agent that enables context-aware conversations about the artworks, triggered through artwork recognition in spatial context.

To make the study interpretable across readers, we implemented WhiteTesseract in a well-known Impressionist setting (Claude Monet works), and use this as a controlled but culturally familiar case. Two assumptions inform our design. First, attentional control, both spatial and perceptual, is a central factor in how visitors engage with artworks, and technological interventions should support rather than disrupt this focus. Second, interpretive engagement benefits from responsive, conversational systems that adapt to the visitor’s level of knowledge, curiosity, and intent, moving beyond the limitations of static labels or one-way information delivery. To examine how these assumptions hold in practice, we developed and deployed WhiteTesseract in a controlled gallery setting using Apple Vision Pro and a curated set of Claude Monet reproductions. We conducted a mixed-methods user study with 26 participants who experienced both traditional White Cube and WhiteTesseract-based visits. Our data include motion tracking, questionnaire responses, system interaction logs, and semi-structured interviews.

The system integrates two complementary mechanisms, an XR-based attention design that modulates visual focus and an AI-powered dialogue that supports reflective interpretation. Rather than isolating these factors, the present study examines their combined influence on visitors’ behavioral and interpretive engagement. Specifically, we ask:
\begin{itemize}
\item \textbf{RQ1}: How does WhiteTesseract system influence visitors’ viewing behavior, particularly in terms of duration and movement patterns, compared to traditional exhibition formats?
\item \textbf{RQ2}: How does the integration of AI-powered dialogue affect visitors’ interpretive engagement with artworks, including the types of questions asked and the perceived depth of interaction?
\end{itemize}
Rather than proposing a radical transformation of gallery spaces, we aim to explore how emerging technologies might enhance key dimensions of the physical cultural heritage experience—namely, attentional focus and interpretive depth—without undermining its curatorial intent. This study contributes to ongoing discussions in HCI and museum interaction design on the role of XR and AI in shaping future exhibition practices grounded in both technological capability and experiential sensitivity.

\section{RELATED WORK}

In physical cultural heritage exhibitions, the viewer’s experience is shaped by the artworks themselves, while by spatial, social, and personal contexts that influence perception and meaning-making \cite{dierking1992redefining}. Although minimalist gallery formats, most notably the White Cube, aim to reduce distractions and promote contemplative viewing, they offer limited tools for adapting to individual differences in perceptual focus or interpretive needs \cite{o1999inside, o2007studio}. At the same time, as digital technologies become increasingly embedded in exhibition settings, new opportunities and challenges arise: how can such tools enhance, rather than disrupt, the viewer’s ability to focus, reflect, and engage \cite{vom2005accounting}?

Prior studies have explored various technologies designed to enhance the visitor experience, including augmented reality overlays, immersive installations, and AI-powered guides \cite{damala2008bridging, tom2017value, rosemary2025leveraging}. However, these systems often address either perceptual or cognitive layers in isolation \cite{chun2007interactions}. Some prioritize immersion or information access, while others seek to scaffold interpretation through text or audio, but without coordinating these functions within the real-time, real-space dynamics of viewing. To better support sustained engagement with artworks, we argue for a combined perspective: one that considers both attentional regulation (e.g., managing sensory input in complex settings) and cognitive support (e.g., enabling personal meaning-making through dialogue or adaptive guidance).

In what follows, we examine three strands of related work that inform our system design: (1) prior applications of XR and AI in exhibition contexts, (2) the aesthetic and perceptual logic of the White Cube model, and (3) the role of conversational AI in shaping reflective interpretation. Together, these domains help frame the design space in which WhiteTesseract operates.

\subsection{XR and AI Technologies in Exhibition Contexts}

XR technologies have been widely adopted in cultural institutions as tools for enhancing presence, accessibility, and immersion \cite{bekele2018survey} Foundational studies in XR emphasize the importance of presence, or the sense of “being there,” and embodiment, which relates to how sensorimotor alignment affects engagement and memory \cite{slater1997framework, biocca1997cyborg}. In museum settings, AR and MR have been used to reconstruct historical scenes, embed digital overlays, or provide layered interpretation through mobile guides \cite{damala2008bridging, styliani2009virtual}.

Yet, such implementations often emphasize content augmentation without adequately addressing perceptual load or attentional coherence \cite{makransky2021cognitive}. Research on DR has shown how selectively removing visual elements can reduce cognitive interference and enhance focus \cite{lee2025diminishar}. Similarly, recent work on passthrough AR highlights how spatially integrated digital content, when precisely aligned with real-world objects, can support seamless transitions between physical and digital modes. These studies suggest that XR's value lies not just in immersion but in the ability to control sensory input—including soundscapes, which are often overlooked \cite{kailas2021design, serafin2020reflections}. For example, XR systems that dynamically manage ambient sound via spatial audio or noise cancellation can improve task performance and subjective focus.

In parallel, museums have begun experimenting with AI systems to support curatorial tasks and enhance visitor engagement \cite{gaia2019engaging, caramiaux2023ai}. Applications range from chatbot-based museum guides to robotic tour companions to personalized recommendation systems \cite{hashemi2018exploiting}. More recently, large language models (LLMs) have enabled flexible, open-ended interactions \cite{naveed2023comprehensive}. For instance, the Smithsonian’s Open Access project integrates GPT-based models for artifact storytelling \cite{bhartiuseful}, and Europe's digital cultural library, museum, and archive has used conversational agents for archival document exploration \cite{machidon2020culturalerica}. However, concerns persist about factual accuracy, hallucination risks in historical contexts, and ethical alignment—especially when LLMs are used in educational or heritage settings \cite{bender2018data, gabriel2023socially}.

Unlike many prior systems that treat XR and AI as separate modalities, visual and textual, respectively, our work explores their coordination for regulating sensory input and interpretive depth \cite{billinghurst2015survey}. WhiteTesseract combines diminished reality with spatial audio and an LLM-based companion that responds to recognized artworks in spatial context, contributing to a hybrid attentional–cognitive environment.

\subsection{Attention, Space, and the Legacy of the White Cube}

The “White Cube” has shaped modern exhibition design for over a century, promoting an environment of sensory minimalism and visual isolation \cite{o1999inside}. As Brian O’Doherty articulated in his seminal critique, the White Cube enforces a purified aesthetic that positions the artwork as autonomous, timeless, and abstracted from social and political contexts. Subsequent critiques have examined this model through broader lenses. Boris Groys argues that the White Cube acts as a frame for power and institutional neutrality \cite{groys2008art}. At the same time, postcolonial perspectives critique its role in reproducing Western epistemologies and excluding diverse forms of engagement \cite{rogoff2013terra, bhambra2014postcolonial, lonetree2012decolonizing}.

From a perceptual perspective, the White Cube is often interpreted as a form of attention design—structuring the viewer’s sensory field to reduce distractions and foster contemplation \cite{bennett2013birth, wood2006attention}. However, cognitive psychology suggests that selective attention is rarely stable or context-free. Classic models of attention demonstrate that perception is shaped by salience, prior knowledge, and task framing \cite{treisman1980feature, posner1980orienting}. In museum contexts, researchers have found that divided attention, resulting from spatial clutter, noise, or crowding, reduces both memory recall and emotional impact \cite{brieber2015white}.

Moreover, recent work in embodied cognition suggests that understanding heritage objects is not purely visual or abstract but mediated through bodily interaction, spatial movement, and multisensory engagement \cite{noe2004action, gibbs2005embodiment}. Eye-tracking studies show that viewing time correlates with interpretive depth and that allowing visitors to adjust their perceptual frame, visually or auditorily, can support more focused exploration \cite{garbutt2020embodied}.

Informed by these traditions, our work treats the White Cube as an architectural reference and a design constraint and opportunity. WhiteTesseract extends this model into a programmable space of attentional modulation, allowing visitors to suppress distractions while preserving the contemplative integrity of minimalist exhibition design \cite{greenberg2005vital, tzortzi2017rethinking}. It also invites inquiry into how spatial structures can be made more responsive to diverse cognitive styles and interaction rhythms.

\subsection{Conversational AI and Dialogic Engagement with Heritage Objects}

Curatorial texts, guided tours, and audio commentaries often scaffold interpretive engagement in museums \cite{falk2018learning, hooper2020museums}. However, each method presents distinct limitations. Curatorial texts offer useful yet fixed contextual information without dynamically responding to individual visitors' interests. Guided tours enable interactive flexibility, but personalized interactions are practically constrained by factors such as the high cost of hiring private guides and limitations in guides' specific expertise. Audio commentaries, although widely accessible, typically follow predetermined sequences and do not support spontaneous questioning or dialogue, limiting adaptability to visitors' unique interpretive needs. In contrast, dialogue-based learning has been shown to improve comprehension, retention, and reflective thinking in educational psychology \cite{chi2001learning, vanlehn2011relative, ruan2019quizbot}. In cultural settings, research indicates that even rhetorical questioning can encourage visitors to spend more time, formulate hypotheses, and form emotional connections with artworks \cite{leinhardt2003learning}.

The advent of conversational AI opens new opportunities for LLMs in art education platforms, enabling open-ended interactions about techniques, styles, and historical movements \cite{mctear2022conversational}. For example, "ArtEMmis" uses vision-language models to generate emotionally resonant descriptions of artworks \cite{achlioptas2021artemis}. At the same time, multimodal systems like "PaLI" combine image input with textual output for captioning and answering questions \cite{chen2022pali}. These models enable personalized and context-aware interpretation \cite{shuster2022blenderbot, kenthapadi2024grounding}, although challenges remain in ensuring factual grounding, inclusivity, and explainability.

In museum contexts, however, the use of conversational AI is still in its early stages of development. Studies have shown that visitors appreciate the ability to ask spontaneous questions but also feel frustration when responses are generic, repetitive, or disconnected from the immediate context \cite{ho2025enhancing}. Our approach builds on this body of work by integrating LLM-driven dialogue into a situated exhibition scenario, where the agent’s prompts are anchored to the viewer’s gaze and the artwork being examined. Rather than simply answering questions, the system aims to act as an interpretive scaffold or advance organizer \cite{ausubel1969school}, encouraging visitors to reflect, compare, and articulate their interpretations through short conversational turns. This approach contributes to ongoing efforts in cultural heritage computing by demonstrating how conversational AI can support personalized, situated interpretation without replacing the embodied authenticity of physical exhibitions.

In museum contexts, however, the use of conversational AI is still in its early stages. Studies have shown that visitors appreciate the ability to ask spontaneous questions but also feel frustration when responses are generic, repetitive, or disconnected from the immediate context \cite{ho2025enhancing}. Recent explorations have introduced more structured approaches, such as graph-based and multi-chatbot architectures that interlink knowledge across collections, and modular platforms that allow curators to build adaptive, human-centered dialogue systems \cite{varitimiadis2021graph,varitimiadis2020towards}. Other works have examined how conversational agents can facilitate question–answer exchanges, personalized recommendations, or even adaptive learning and game-based engagement in museum contexts \cite{schaffer2021questions,tsitseklis2023recbot,liang2024effect,gaia2019engaging}. While these approaches significantly advance personalization and interactivity, their logic remains largely data-centric, focused on curating and adapting textual responses to user queries. Our approach builds on this body of work by integrating LLM-driven dialogue into a situated exhibition scenario, where the agent’s prompts are anchored to the viewer’s gaze and the artwork being examined. Rather than simply answering questions, the system acts as an interpretive scaffold or advance organizer \cite{ausubel1969school}—encouraging visitors to reflect, compare, and articulate their interpretations through short conversational turns. In this way, WhiteTesseract reframes personalization as an embodied and context-aware process that co-evolves with the visitor’s sensory and spatial experience, demonstrating how conversational AI can support situated interpretation without replacing the authenticity of physical exhibitions.

\section{WHITETESSERACT: EXTENDING WHITE CUBE INTO EXTENDED REALITY}

\subsection{Conceptual Motivation: Beyond the Traditional White Cube}
The WhiteTesseract concept emerged from a critical examination of how high-resolution XR technology could meaningfully extend, rather than replace, the established principles of the White Cube gallery space. The White Cube, as a curatorial paradigm, creates a visually neutral environment intended to foster focused contemplation. Yet, it offers limited adaptability to individual viewers’ perceptual needs or interpretive preferences.
WhiteTesseract reinterprets this model by introducing dynamic, responsive digital elements that augment the physical exhibition space. Rather than treating technology as an add-on, we explore how XR and AI can be integrated to reshape attention, perception, and meaning-making within the gallery experience.

\subsection{System Design: Coordinating Perception and Dialogue}
\begin{figure}[ht]
    \centering
    \includegraphics[width=1\linewidth]{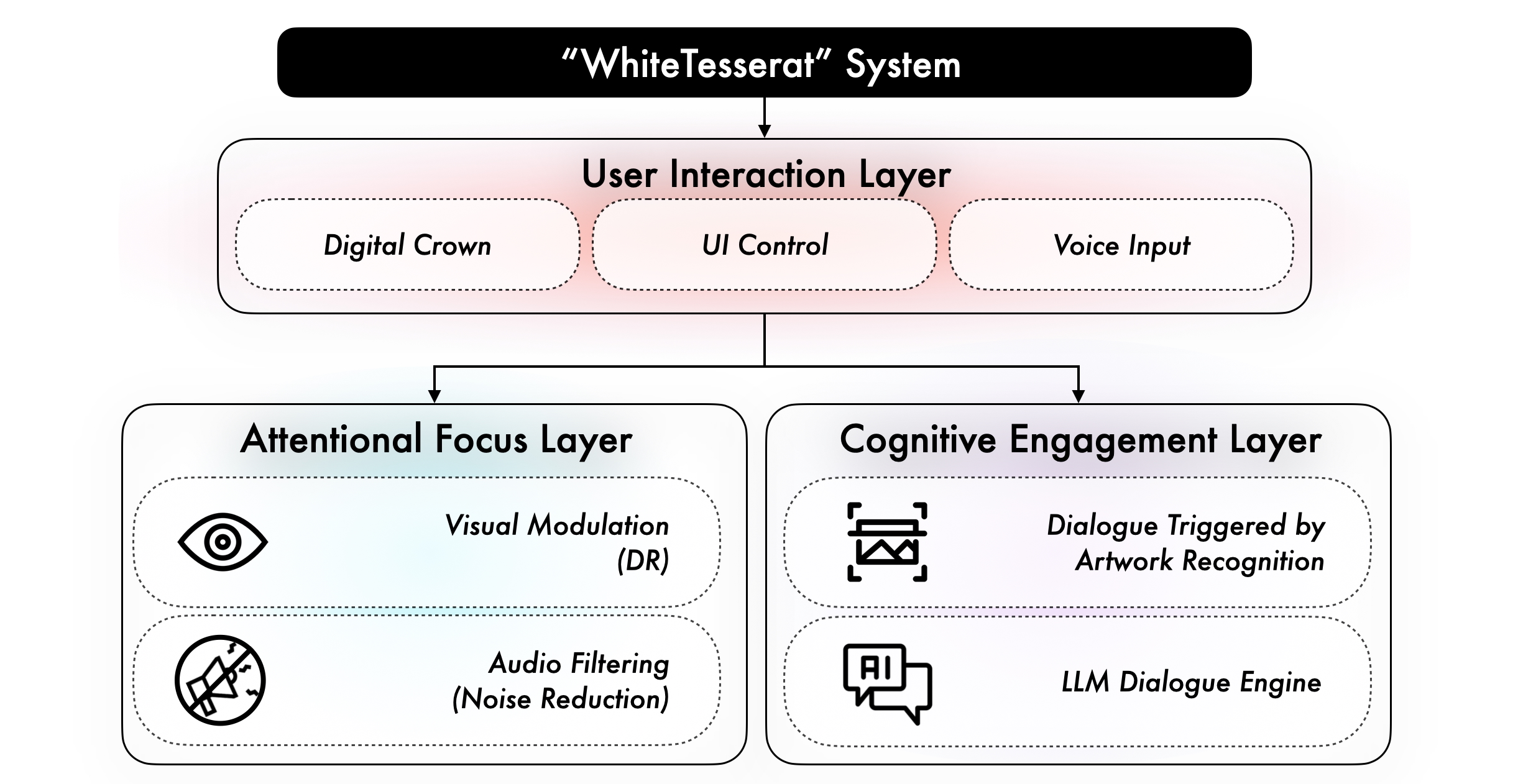}
    \caption{WhiteTesseract Design Framework: Coordinating Attentional Focus And Cognitive Engagement. The system integrates user interaction controls that jointly regulate attention (XR-based) and cognitive engagement (AI-based), forming an adaptive exhibition framework.}
    \label{fig:WhiteTesseract Design Framework}
\end{figure}

WhiteTesseract integrates two interdependent design layers to support visitor engagement with artworks in real-world gallery settings: one centered on attentional focus and the other on cognitive engagement. The attentional layer enables visitors to dynamically reduce sensory distractions, while the cognitive layer facilitates lightweight, situated dialogue. These layers are not isolated modules but are synchronized through spatial awareness and user control, creating a flexible perceptual and interpretive environment. Figure \ref{fig:WhiteTesseract Design Framework} outlines the system’s conceptual framework, highlighting how these components interact to form an adaptive exhibition experience.

\begin{figure}[ht]
    \centering
    \includegraphics[width=1\linewidth]{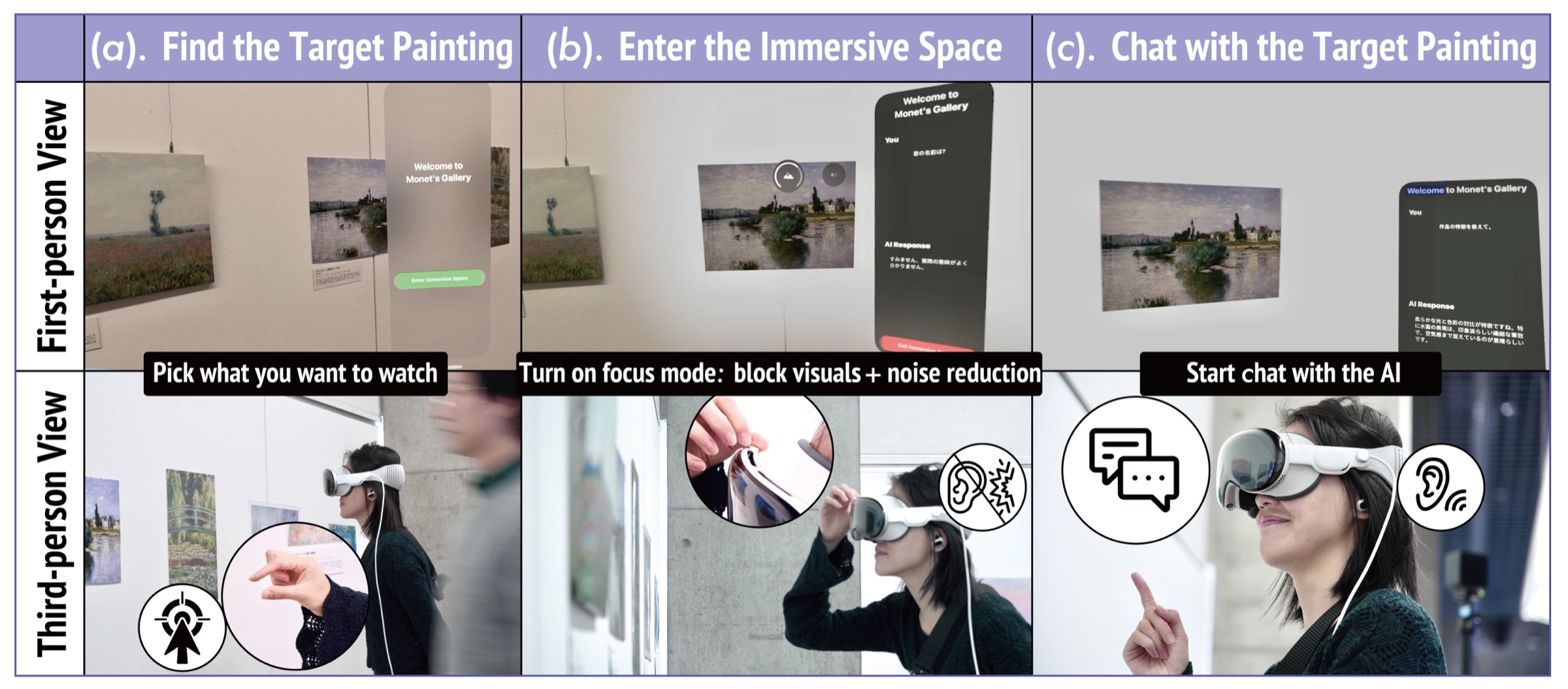}
    \caption{User Interaction Flow in WhiteTesseract. The three-stage journey includes: (a) artwork selection through gaze, (b) entrance into an immersive perceptual space with visual/auditory isolation, and (c) natural language dialogue with the selected artwork. First-person and third-person views are shown to illustrate perceptual adjustment and behavioral interaction.}
    \label{fig:Usage Process}
\end{figure}

To illustrate how these two components unfold in practice, Figure \ref{fig:Usage Process} presents a representative user scenario from both first-person and third-person perspectives. The interaction proceeds through three coordinated phases: \textbf{target identification}, \textbf{perceptual transition}, and \textbf{dialogue initiation}. In the first phase, the user naturally focuses on a painting of interest in the physical gallery. The system detects this target using computer vision-based recognition, which relies on subtle digital labels embedded in the scene. This recognition occurs passively and does not require explicit user confirmation, allowing the gallery to remain visually minimal and uninterrupted. Once the target is identified, the user activates the immersive mode using the headset’s Digital Crown, which gradually overlays a visual mask that blends with the gallery’s background colors. This masking is coupled with spatial audio suppression via in-ear filtering, reducing ambient distractions. Rather than instantly shifting the environment, these perceptual adjustments unfold gradually, enabling the user to focus on the selected painting while minimizing external interference.

After the perceptual field has been adjusted, the system enters a state of visual and auditory isolation around the selected artwork. At this stage, the LLM initiates the interaction by offering a context-sensitive prompt—such as a reflective question or interpretive cue, without requiring the user to issue a command. The user can respond directly via speech, which the system passively captures and interprets in real time. To support accessibility and comprehension, all dialogue is visually represented in a lightweight on-screen chat box that users can reposition or resize as needed. This interface is intentionally unobtrusive and designed to respect the artwork's centrality; it appears only when active and remains under user control. While the system avoids intrusive overlays, the optional visual feedback offers a subtitle-like experience for those who prefer textual confirmation, supporting diverse interaction preferences without compromising visual focus.

This coordinated design transforms how sensory and semantic attention are managed in exhibition contexts. Instead of guiding the user through fixed pathways or scripted content, WhiteTesseract enables a responsive, visitor-driven rhythm of seeing and thinking, consistent with the contemplative intent of the White Cube while expanding its expressive potential.

\subsection{Reframing the Exhibition Experience}
The integration of attentional and cognitive modulation in WhiteTesseract alters how visitors interact with individual artworks, while how they structure time, movement, and reflection within the exhibition space. Traditional gallery design typically assumes a linear and uniform, uniform visitor journey, in which perceptual pacing and interpretive content are space and time. By contrast, WhiteTesseract enables an adaptive exhibition model in which visitors can regulate their sensory environment and dialogic engagement in real-time.

This shift foregrounds the exhibition not as a fixed presentation of objects but as a dynamic perceptual interface, inviting visitors to inhabit the gallery according to their attentional rhythms and interpretive needs. Rather than offering interactivity as spectacle or entertainment, the system facilitates a quieter form of agency: the ability to dwell, to inquire, and to return.

This reconceptualization of the exhibition experience is operationalized in three specific design contributions:

\begin{itemize}
\item \textbf{Decoupling spatial layout from cognitive sequencing}: WhiteTesseract enables visitors to freely choose their own viewing order and interpretive routes, independent of physical arrangement or sequential wall texts. This facilitates flexible exploration, allowing revisits and varied interaction sequences without altering gallery architecture.
\item \textbf{Reframing peripheral attention through environmental modulation}: By enabling visitors to focus or isolate artworks on demand, the system dissolves the hierarchy between central and marginal works, expanding curatorial reach without redesigning space.
\item \textbf{Transforming silence and solitude into configurable conditions of engagement}: Through real-time control of visual and auditory input, visitors can choose when and how to concentrate, turning perceptual quiet into an active interpretive resource—rather than a passive default.
\end{itemize}
Together, these contributions position WhiteTesseract not simply as a technical enhancement of the gallery space, but as a reframing of how exhibitions function as cognitive and attentional systems.

\section{TECHNICAL IMPLEMENTATION}

\begin{figure}[ht]
    \centering
    \includegraphics[width=1\linewidth]{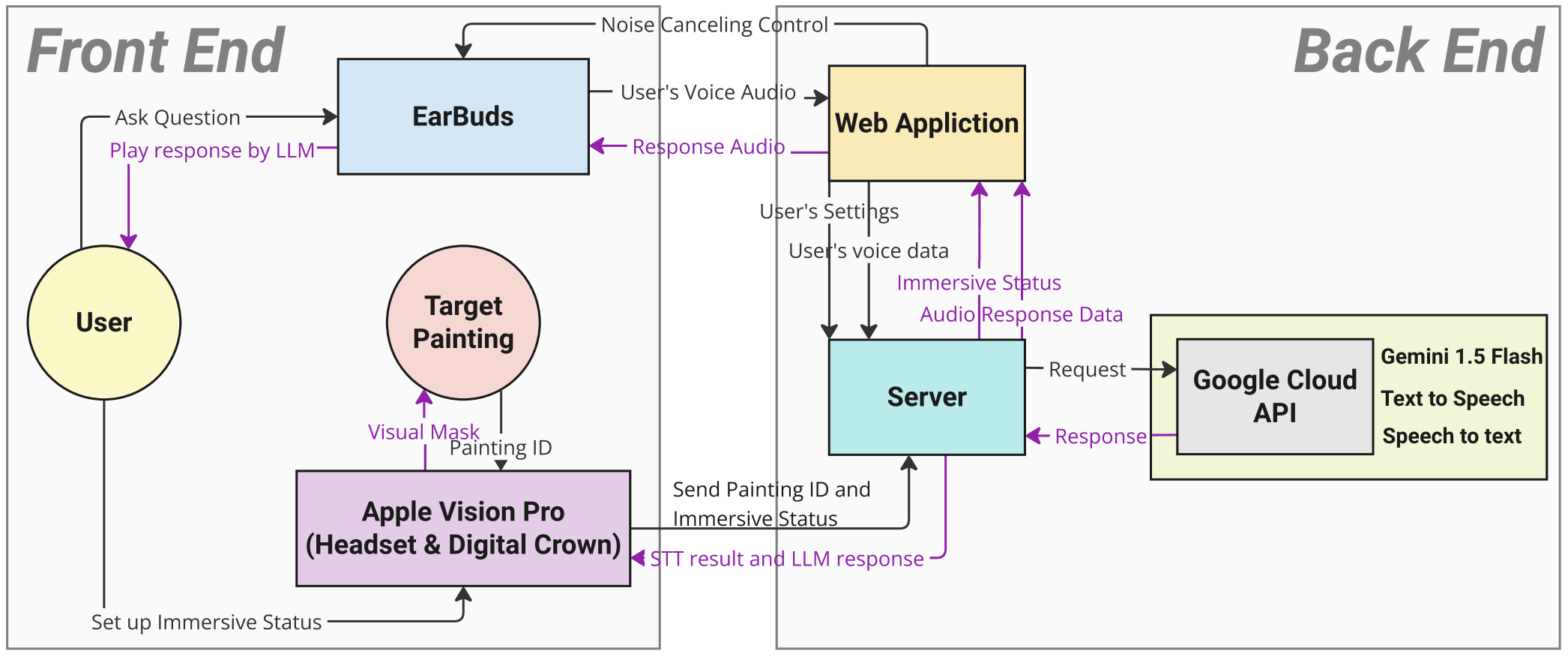}
    \caption{The “WhiteTesseract” system. It includes an Apple Vision Pro, earbuds with noise canceling function and a server is built with Google's LLM Gemini 1.5 Flash, and Speech-to-text/Text-to-speech APIs for handling LLM conversation.}
    \label{fig:system diagram}
\end{figure}

The technical implementation of the \textit{digital reduction} framework integrates XR and LLM to create an optimized environment for art appreciation. This section details the methodologies and technologies employed to achieve distraction reduction and facilitate in-depth communication between visitors and artworks. The “WhiteTesseract” system consists a Apple Vision Pro and a pair of earbuds (Fig. \ref{fig:system diagram}, equipments, 1 and 2), while a motion tracker (Fig. \ref{fig:system diagram}, equipments, 3) is dedicated solely to recording visitor movement data. The whole system's function diagram is demonstrated in Fig. \ref{fig:system diagram}.

\subsection{XR-Reducing Distractions}

To effectively reduce distractions, our system utilizes XR technology to adjust the visibility of surrounding elements dynamically. Utilizing spatial anchors and image recognition, the WhiteTesseract application identifies and selectively occludes visual clutter that is unrelated to the artwork being viewed. The Apple Vision Pro was selected due to its high resolution (3660 × 3200 pixels per eye) and minimal image distortion. The selective passthrough functionality is implemented through VisionOS Tracking, which utilizes preregistered images via the 3D space API\footnote{https://developer.apple.com/documentation/visionos/tracking-images-in-3d-space}. This allows a mask to be spatially anchored and overlaid on the volume view, excluding the recognized artwork, enabling users to selectively block visual clutter by rotating the Digital Crown. We also linked the Digital Crown with the active noise cancellation earbuds (Inzone Buds WF-G700N) to reduce noise. When the user enters the "Immersive Space", the earbuds automatically activate noise cancellation. These designs aim to provide an "isolated environment" that enhances the user experience through a digital approach, blocking real-world interference.

\subsection{LLM - Establishing in-depth Communication}

The dialogue system is integrated with an LLM to foster meaningful interactions between visitors and artworks. We developed the LLM conversation system combined with Apple Vision pro's graphic interface with Google's Cloud APIs. The system utilized Gemini 1.5 Flash \cite{gemini_flash} for constructing conversation experience. Considering our goal was to use LLM for providing more background knowledge of the artworks, and this didn't require a complex reasoning function, the API we used here is the "generate content" API that doesn't store any history data. The user's voice recorded by the earbuds is transcribed through Google's cloud Speech-to-text (STT) API\footnote{https://cloud.google.com/speech-to-text?hl=en}, and the text content is sent to the LLM for a response.  Google's cloud Text-to-speech (TTS) API\footnote{https://cloud.google.com/text-to-speech?hl=en} will render the reply into speech audio.  For a better overview, we set up a server in the local network and built a web-based application for experiments. The server handles all API operations and sends and receives the artwork ID from Apple Vision Pro's viewport recognition function by HTTP Post.  The web application is used to set the dialogue system's language, control the earbuds' noise-canceling state, and record conversation content. The entire dialogue process is shown in Fig.\ref{fig:system diagram}. 

We expected the LLM to react like a humanized agent in the immersive space, evoking the user's deep understanding of the artwork. However, the LLM program has limitations in understanding users' visit behavior, which may affect users' appreciation experience. Suitable prompt patterns can help conquer this problem \cite{promptPattern}. We set up specific prompt patterns include \textit{language code} and \textit{Painting name} to ensure the LLM can respond to the user's question properly in the exhibition scenario and use the correct language. The \textit{language code} is the preload parameter selected by the user in the web application. The \textit{Painting name} is dynamically changed according to the recognition result provided by Apple Vision Pro. The complete prompt except user's question is:

 \textit{"There's a Claude Monet exhibition going on right now. Now the user is watching the painting: (\textbf{Painting Name}). Please act like an accompanying person in this exhibition, briefly answer the following question with valuable knowledge and information, not more than 40 words, and use language in (\textbf{language-code}). Attention: The user may have already asked you a question about the painting. If the painting is not changed, please do not repeat the same information, especially the painting's name. If the user's question/expression is not related to the painting or is confusing, please politely say something like 'I'm sorry, I don't know how to answer this question.' use language in (\textbf{language-code})}."

We designed the conversation system into a form of alternating questions and answers: The voice detection algorithm will send the STT result combined with the prompt to Gemini 1.5 Flash if there is a 3 seconds silence after the user's speaking. After each triggered input, the LLM (Gemini 1.5 Flash) used its default generation parameters, which provided a balanced level of response variability suitable for conversational tasks. Then the voice synthesis program will use the TTS API to generate audio after the LLM has finished its response, which may have a delay of around 2 to 5 seconds, according to the API's condition. Through repeated testing, we ensured that the total response delay can be controlled under 8 seconds. During the voice responding period, the speaking detection algorithm will keep pending. The user cannot interrupt this process while the LLM is "speaking". But there may be situations where APIs don't respond within 8 seconds or just return error messages. In such cases, the program will immediately exit the current pending conversation and start a new one. The user is notified of the current conversation state by the visual hint from Apple Vision Pro. When the LLM API works normally, the response text is displayed in the conversation GUI; otherwise, the interface indicates that the conversation is available.

\section{USER STUDY}

We chose a Claude Monet exhibition as the testbed because it represents a widely known yet often passively consumed body of visual heritage, allowing us to observe whether XR + AI could deepen engagement even in a familiar setting. As a foundational figure of the Impressionist movement, Monet's practice not only transformed modern visual culture but also established a legacy that continues to influence contemporary understandings of light, atmosphere, and perception. While his paintings are aesthetically familiar and widely appreciated, they often evoke limited critical engagement in traditional gallery settings, where visitors may view primarily for affect or enjoyment rather than deeper analysis \cite{bennett1995birth, hooper2020museums, darda2025comparison}. This choice allowed us to explore how perceptual modulation and dialogue-based interpretation could enrich visitors’ attention and understanding in a controlled yet culturally significant scenario. The user study took place in a real gallery environment using reproductions of Monet's works, enabling in-situ evaluation of WhiteTesseract’s effects on cultural engagement.

\subsection{Participants and Setting}

To evaluate our design system and explore potential insights and design space, we conducted a user study in the local citizen art gallery. The user study was approved by the University of Aaaaaaa's ethics committee. The entire study lasted 5 days; 27 participants, recruited from the university student community, visited the gallery. However, due to the malfunction of the motion tracker, the tracking data for the first participant was incomplete. Due to the gallery's time restrictions, the Number 16 participant was unable to experience the entire experiment. As a result, only 26 participants who completed the entire experiment were included in the final analysis. The final valid amount of questionnaire data collected was 26 participants, they reported their ages within the provided ranges: 20–25 (N=12, 46\%), 25–30 (N=14, 54\%). The amount of tracking data collected before using our system was 25, while the amount of tracking data collected after using our system was 26. Detailed demographic attributes, including participants’ age, gender, academic background, familiarity with XR technologies, and expectations for AR-enhanced art experiences—are summarized in Table~\ref{tab:participant_demographics} in Appendix~A.

\begin{figure}[ht]
    \centering
    \includegraphics[width=1\linewidth]{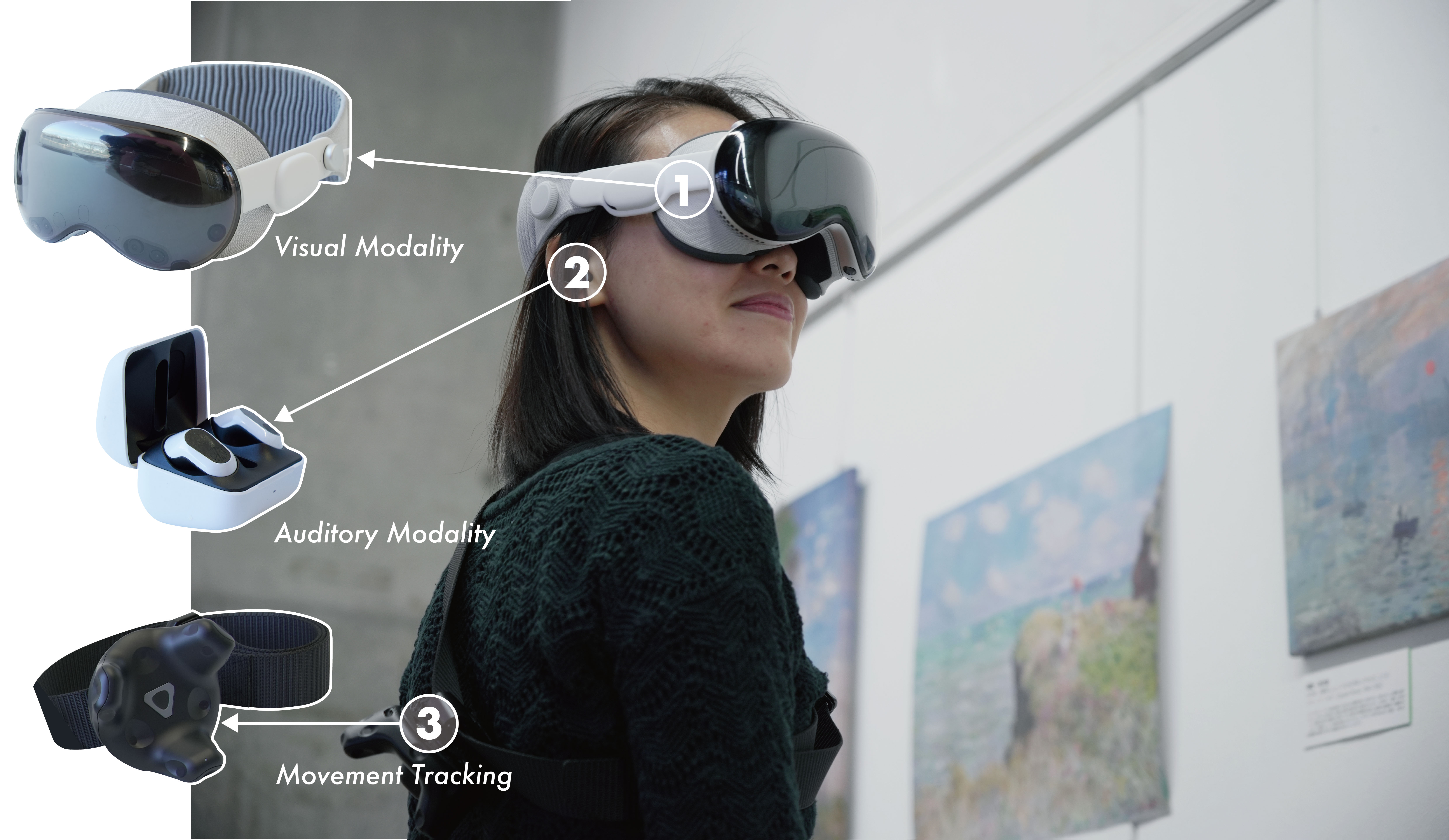}
    \caption{Equipment Configuration of Participant During User Study}
    \label{fig:Equipment}
\end{figure}

As demonstrated in Fig. \ref{fig:Experimental Scenario Design}, we provide a detailed view of the experimental scenario design. Panel (a) illustrates the two-dimensional layout of the visitor activities area, indicating precise dimensions (5m by 4.5m) and clearly marking the designated starting point for participants. In this paper, each artwork is assigned a Painting ID (No. 1–No. 10), which serves as a system-level identifier corresponding to a fixed physical location in the exhibition space. Painting IDs are used consistently across motion tracking, viewing-duration analysis, and AI dialogue logs, and are mapped to artwork titles as metadata rather than representing the titles themselves. Panel (b) visually presents the physical setup of the exhibition space, showing ten reproductions of Claude Monet’s artworks arranged systematically along the walls. Detailed information regarding the artworks, including titles and dimensions, is summarized in accompanying tables. Panel (c) depicts a participant actively engaging with the artworks within the defined space, demonstrating the practical implementation of the experimental scenario. The spatial layout, artwork order, print sizes, and lighting were identical across both the traditional and WhiteTesseract system conditions to ensure a like-for-like comparison. Fig. \ref{fig:Equipment} illustrates the equipment configuration used by participants during the user study. It highlights three primary technological components integral to the WhiteTesseract system: (1) visual modality provided by the Apple Vision Pro headset, which delivers immersive high-resolution visuals; (2) auditory modality managed through wireless earbuds, enabling controlled audio feedback and ambient sound suppression; and (3) movement tracking facilitated by a dedicated tracking device secured to the participant, capturing precise behavioral and movement data for subsequent detailed analytics.

\begin{figure}[ht]
    \centering
    \includegraphics[width=1\linewidth]{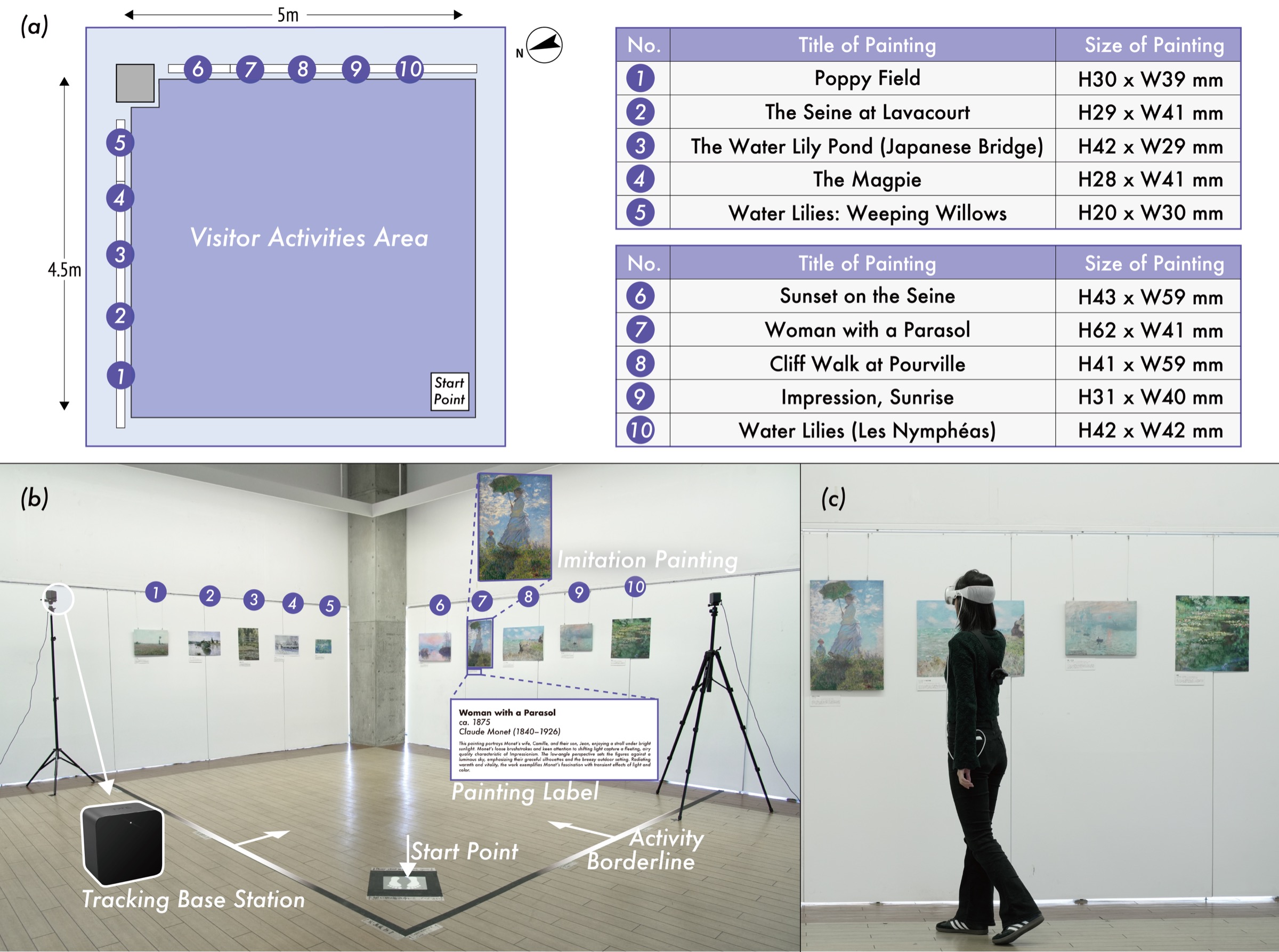}
    \caption{Experimental Scenario Design. (a) is the actual visiting area in 2D. The “No.” denotes the Painting ID, a system-level identifier linked to a fixed artwork location in the exhibition and mapped to the corresponding artwork title. As shown in (b) we have set up a specific visiting place featuring 10 different artworks by Claude Monet. The information about the paintings is listed in the tables. Participants can visit any artwork they wish (c) in the designated area.}
    \label{fig:Experimental Scenario Design}
\end{figure}

Participants wore a lightweight multi-modal setup during the experiment (see fig. \ref{fig:Equipment}), consisting of a high-resolution XR headset (Apple Vision Pro), in-ear spatial audio devices, and a body-mounted movement tracker. This configuration supported immersive visual masking, gradual ambient sound suppression, and fine-grained motion tracking throughout the exhibition. The integrated setup enabled both perceptual modulation and conversational interaction while capturing behavioral data, including head orientation and viewing duration.

\subsection{Study Procedure}

The study followed a within-subjects design comparing traditional and WhiteTesseract conditions. In the XR condition, both the diminished reality feature and the conversational AI were active, allowing us to evaluate their combined influence on art-viewing behavior. Each session lasted approximately 75 to 100 minutes, beginning with an orientation and informed consent, followed by a pre-experience survey to gather demographic data and initial impressions.

\begin{figure}[H]
    \centering
    \includegraphics[width=1\linewidth]{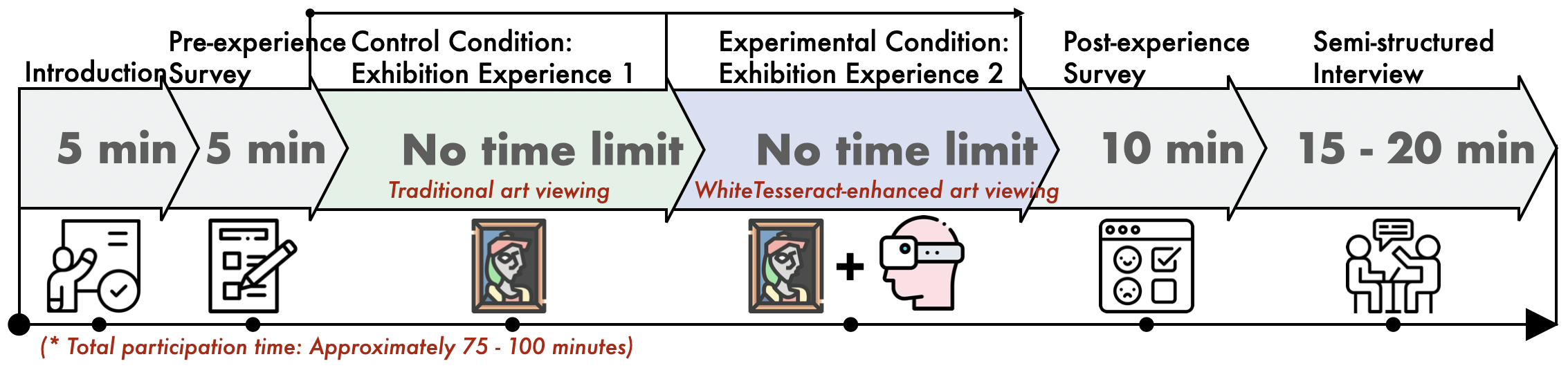}
    \caption{User study procedure. Each participant will experience traditional visiting first, then use the XR-enhanced system to visit the gallery again.}
    \label{fig:user}
\end{figure}

Participants then engaged in two distinct viewing experiences: one using traditional gallery methods and the other employing the WhiteTesseract XR system. In the traditional viewing condition, participants experienced the same 5 × 4.5 m layout and artwork sequence as in the WhiteTesseract system condition, but without any headset or audio devices. Ten Monet reproductions were printed at identical sizes and mounted at eye level under uniform gallery lighting. Participants received a brief written instruction, “Please view freely as in a normal exhibition,” and could move naturally without time limits. Only physical wall labels were available (title, date, medium); no conversational AI or audio suppression was provided. Following these viewing experiences, participants completed a post-experience survey assessing their reactions to both conditions. Subsequently, they participated in semi-structured interviews lasting approximately 15 to 20 minutes to capture detailed subjective feedback. The detailed procedure flow is illustrated in Fig. \ref{fig:user}.

\subsection{Data Collection and Analysis}

To systematically address our research questions, we employed multiple data collection methods and corresponding analytical approaches tailored to examine the overall behavioral effects of the WhiteTesseract system (RQ1) and AI-facilitated dialogue (RQ2).

\subsubsection{Motion Tracking and Behavioral Analysis (RQ1)}

Motion tracking data were collected using HTC Vive Tracker (first generation) and base stations to capture participants' moving trajectory throughout the gallery. The positions of artworks served as spatial references to define 10 viewing areas. For detailed information about the artworks' positions in the physical exhibition space, please refer to panels (a) and (b) in Fig. \ref{fig:Experimental Scenario Design}. Movement trajectories were projected onto a top-view to calculate viewing duration for each artwork. Since participants could revisit artworks multiple times, we cumulatively summed the time spent during each visit to obtain total viewing duration per area. Entry and exit points were determined using predefined polygonal boundaries around each artwork. Statistical comparisons between the control condition (traditional viewing) and experimental condition (WhiteTesseract) employed Mann-Whitney U tests with significance level set at \textit{p} < 0.05. Importantly, the area definitions and timing rules (entry/exit thresholds, revisits) were applied identically in both the traditional and WhiteTesseract system conditions.

\subsubsection{Conversation Analysis and Interview Coding (RQ2)}
All 529 visitor–AI conversations were collected from system logs and translated from Japanese and Chinese to English by bilingual researchers.
To establish a clear, theoretically grounded classification scheme, three researchers conducted open coding on a random sample of 100 questions to identify recurrent intents and linguistic patterns. This inductive analysis was informed by prior research on visitor questioning and interpretive dialogue in museum learning contexts \cite{leinhardt2003learning, serrell2015exhibit, bitgood2016attention}. Based on these foundations, six categories were defined and refined through iterative discussion: Descriptive Information, Analytical Understanding, Emotional and Personal Engagement, Comparative and Contextual Inquiry, Technical and Functional Questions, and Open-Ended Exploration (Table \ref{tab:catalogs of questions}). After consensus on the definitions was reached, two researchers independently coded the entire dataset, and discrepancies were resolved in group sessions to ensure consistency and reliability. 

To analyze patterns in AI responses, we employed the Natural Language Toolkit (NLTK\footnote{https://www.nltk.org/index.html}) to extract word and phrase frequencies from the translated responses. The top 20 most frequent terms were visualized using word clouds to identify dominant themes and potential repetitiveness in the AI's vocabulary.

Questionnaire data included the User Experience Questionnaire (UEQ) for evaluating eight dimensions of user experience \cite{laugwitz2008construction}. The specific adjective pairs used in the adapted UEQ—such as obstructive–supportive, confusing–clear, and conventional–inventive, are listed in Table~\ref{tab:ueq} (Appendix~B). Additionally, 16 custom Likert-scale items were designed to address AR technology acceptance and AI dialogue satisfaction, including participants’ perceptions of immersion, visual clarity, comfort, learning value, and adaptability of the voice-based guide. The full list of AR-related and LLM-related items is provided in Table~\ref{tab:post_experiment_questionnaire} and Table~\ref{tab:llm_questionnaire}, respectively. Spearman's rank correlation coefficient was calculated between UEQ total scores and individual custom questionnaire items to identify key factors influencing overall user experience.

Semi-structured interviews lasting 15–20 minutes were conducted with all 26 participants immediately following their experimental sessions. Interviews were audio-recorded, transcribed verbatim, and translated into English by certified translators. The interview protocol included prompts targeting participants’ impressions of the XR system, interaction with the AI guide, and comparisons with traditional exhibitions. The full set of interview questions is listed in Table~\ref{tab:Semi-structured} (Appendix~C). Interview data underwent systematic thematic analysis following established qualitative research protocols, where three researchers independently reviewed the transcripts to identify initial themes and patterns, with particular attention to: (1) visual immersive experience with the XR system; (2) conversational interaction experience with the AI; (3) perceived benefits and limitations of the system; and (4) comparisons with traditional museum experiences. Following individual analysis, the research team convened for collaborative sessions to compare emergent themes, reconcile differences in interpretation, and develop a unified coding framework. This iterative process of independent analysis followed by group discussion ensured both analytical rigor and interpretive depth, allowing for nuanced understanding of participants' experiential journeys through the WhiteTesseract system. The final thematic structure emerged through consensus, capturing both convergent patterns and divergent perspectives across participants, who are referenced using anonymous identifiers (P1–P27) throughout the analysis to maintain confidentiality while preserving individual voices.

\begin{table}[H]
    \centering
    \caption{Catalogs of participants' questions in LLM conversations}
    \small
    \begin{tabular}{p{4.5cm} p{4.5cm} p{5cm}} 
        \hline
        \textbf{Catalog} & \textbf{Definition} & \textbf{Examples} \\ \hline
        Descriptive Information & 
        Questions about the basic facts, background, or historical information of the Artwork & 
        What is the name of this work? \newline Where was this painted? \\ \hline
        Analytical Understanding & 
        Questions about the artistic technique, meaning, and style of the Artwork & 
        What technique was used to paint this? \newline What is the main focus of this painting? \\ \hline
        Emotional and Personal Engagement & 
        Questions, expressions or emotional reactions to the Artwork & 
        Why does Monet like water lilies? \newline The depiction is beautiful! \\ \hline
        Comparative and Contextual Inquiry & 
        Questions about other related artworks, artists, or cultures & 
        How many other paintings of the Seine besides this one? \newline Did Monet's painting style change throughout his career? \\ \hline
        Technical and Functional Questions & 
        Questions about immersive space systems or other technologies & 
        Can I zoom out on this painting? \newline Can you show me this place on Google Maps? \\ \hline
        Open-Ended Exploration & 
        Divergent questions reflecting creative/deep thinking of the artwork & 
        Why was Monet fascinated by Japanese culture? \newline What are the characteristics of the wind movements? \\ \hline
    \end{tabular}
    \label{tab:catalogs of questions}
\end{table}

\section{RESULTS}

\subsection{Qualitative Results}

\subsubsection{Questions about the artworks}

\begin{figure}[!hb]
    \centering
    \includegraphics[width=1.0\linewidth]{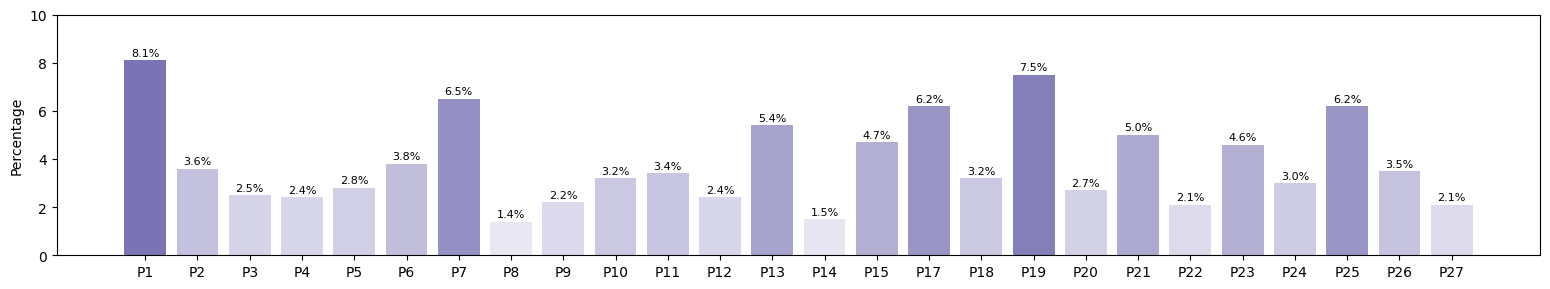}
    \caption{isualization of the proportion of each participant’s question quantity based on a total of 529 visitor-generated questions.}
    \raggedright
\footnotesize\textit{Note: Questions associated with Artwork No.~2 were not included due to an initial labeling mismatch between two visually similar artworks (The Seine at Lavacourt and The Seine near Giverny). In addition, data from Participant 16 (P16) were excluded due to incomplete participation.}
    \label{fig:ques_each_distribution}
\end{figure}

The classification of question types in Table~\ref{tab:question_distribution}, specifically whether they are typically addressed in traditional museum labels, is based on a comparative analysis of curatorial practices at leading institutions such as the Museum of Modern Art (MoMA) and the National Gallery (UK), along with findings from exhibition design literature \cite{serrell2015exhibit, leinhardt2004listening, bitgood2016attention}. Prior research indicates that conventional wall texts prioritize factual, objective content—artist name, artwork title, medium, date, and brief contextual notes—while deliberately limiting subjective, speculative, or interpretive elaboration. These practices reflect curatorial norms that emphasize clarity, neutrality, and accessibility over emotional engagement or dialogic provocation. Against this backdrop, our analysis of 529 visitor-generated questions during AI-mediated art viewing revealed markedly different patterns of interpretive engagement. The proportions of the quantity of each participant's questions are demonstrated in Fig. \ref{fig:ques_each_distribution} As shown in Table~\ref{tab:question_distribution}, while 40.8\% of the questions focused on descriptive information (in line with what traditional labels typically provide), the remaining 59.2\% extended beyond conventional scope—encompassing analytical, emotional, comparative, technical, and open-ended inquiries. This contrast highlights the expanded interpretive space enabled by conversational AI systems, which invite forms of inquiry rarely supported by standard museum texts.

The statistical results for each painting (Fig. \ref{fig:ques_response}) show that while \textit{descriptive information} and \textit{analytical understanding} formed the majority of inquiries, the nature of these questions revealed unexpected depth. Within the descriptive category, participants pursued highly specific details rarely covered in standard museum texts. For instance, one participant asked about "Impression, Sunrise": "\textit{Which port did he paint?}"—seeking precise geographical information beyond typical label content. Another examining "Sunset on the Seine" inquired: "\textit{It seems like there's someone rowing a boat in the middle, what is this?}"—demonstrating close visual attention to compositional details.
Perhaps most revealing was participants' interest in human narratives behind the artworks. Questions such as "\textit{Did the model of the child, his wife Camille, and son Johnny have to stand there the whole time while painting?}" for "Woman with a Parasol" indicate that the conversational format encouraged visitors to consider the lived experiences of artistic creation. This shift from viewing artworks as static objects to understanding them as products of human processes suggests that AI dialogue can fundamentally alter the interpretive framework visitors bring to museum experiences.

\begin{table}[ht]
\centering
\caption{Distribution of Question Types in AI-Mediated Dialogue vs. Traditional Museum Labels}
\begin{tabularx}{\textwidth}{@{}p{0.32\textwidth} c c p{0.52\textwidth}@{}}
\toprule
\textbf{Question Type} & \textbf{N} & \textbf{\%} & \textbf{Typically Provided in Traditional Labels} \\
\midrule
Descriptive Information & 219 & 41.4 & Yes – Primary focus (e.g., title, date, medium, dimensions, basic historical context) \\
Analytical Understanding & 100 & 18.9 & Limited – Occasionally includes artistic technique or style notes \\
Emotional and Personal Engagement & 56 & 10.6 & No – Rarely addresses personal or emotional dimensions \\
Comparative and Contextual Inquiry & 64 & 12.1 & Limited – May reference artistic movement but rarely compares works \\
Technical and Functional Questions & 34 & 6.4 & N/A – Not applicable to traditional labels \\
Open-Ended Exploration & 56 & 10.6 & No – Does not support speculative or creative inquiry \\
\midrule
\textbf{Total} & \textbf{529} & \textbf{100} & \\
\bottomrule
\end{tabularx}
\label{tab:question_distribution}
\raggedright
\footnotesize\textit{Note: Classification based on AI-dialogue content analysis; comparison with conventional museum label formats derived from exhibition catalog review. Questions associated with Artwork No.~2 were not included in this analysis due to an initial labeling mismatch between two visually similar artworks (The Seine at Lavacourt and The Seine near Giverny).}
\end{table}

\subsubsection{LLM's Responses}
\begin{figure}[ht]
    \centering
    \includegraphics[width=1.0\linewidth]{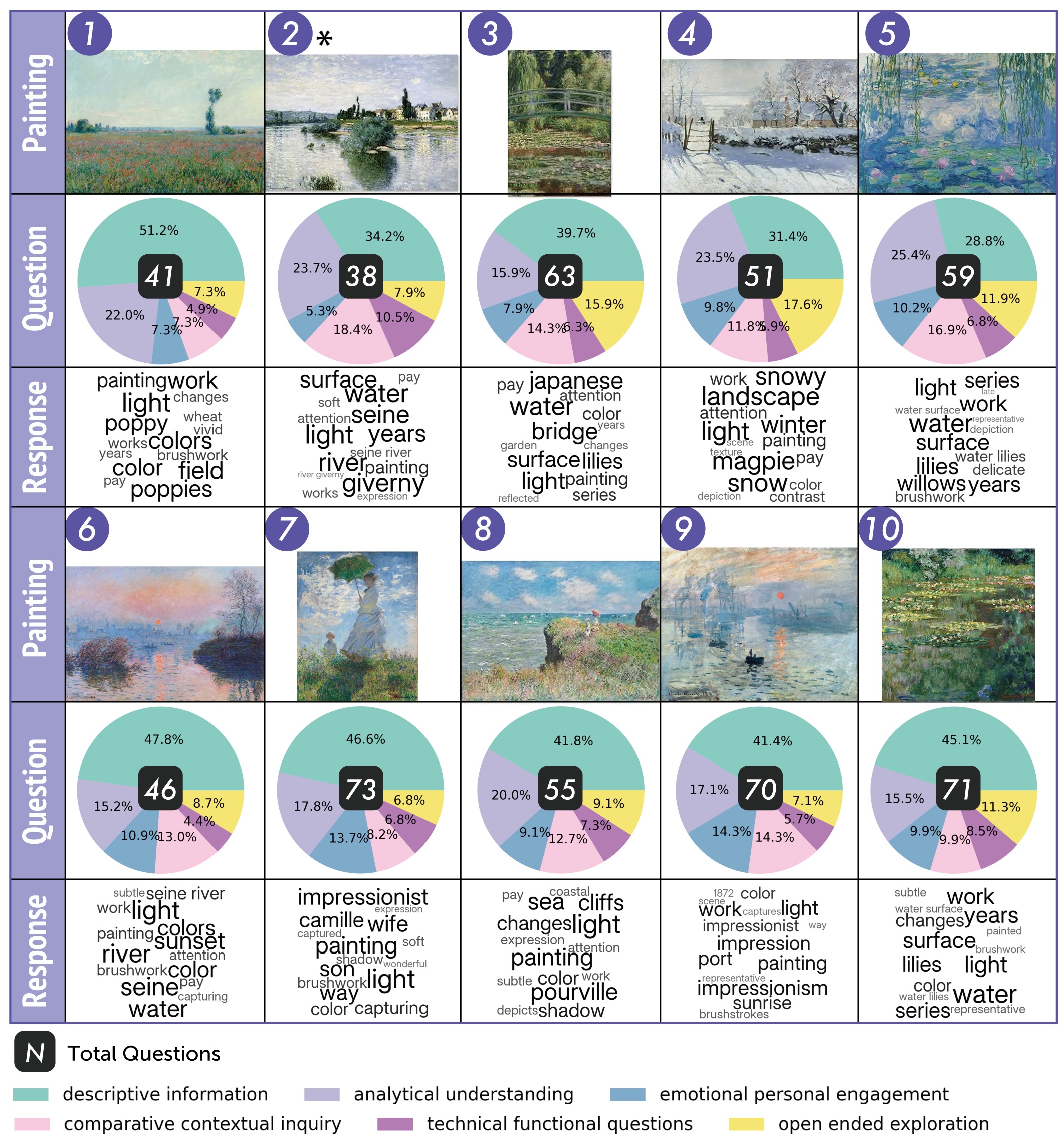}
    \caption{Visualization of participants' questions in the LLM dialogue system and the word cloud of LLM's responses. The "No." refers to the artwork number in the gallery, the Pie charts show each catalog's proportion of all questions, and the word cloud in the response rows demonstrates the high-frequency response words from LLM.}
    \vspace{2pt}
\raggedright
\footnotesize\textit{Note: * indicates an initial labeling mismatch between two visually similar artworks (The Seine at Lavacourt and The Seine near Giverny). Due to this ambiguity, the corresponding LLM responses were not considered analytically interpretable for comparative analysis.}
    \label{fig:ques_response}
\end{figure}

Based on the results of the frequency analysis of words and phrases frequency analysis results (Fig. \ref{fig:ques_response}), the responses from LLM seem repetitive and a little rigid. Imposing restrictions on prompts will cause some negative effects on conversational interaction. "\textit{Light}" and "\textit{Color}" - the core concepts in Impression artworks - have been mentioned in a continuous high frequency, which might confuse participants when they want more distinctive knowledge. But we were also impressed by some inspirational responses that actually had the potential to promote more in-depth cognition of art. One participant asked about the artwork "Poppy Field": "\textit{Does Monet's work mainly depict dynamic scenes?}". The LLM's response was: "\textit{Not entirely. Monet's works depict both dynamic light and shadow, such as the swaying poppies in this work, and relatively static scenes, like the Water Lilies series. He focused more on the momentary changes of light and color.}" Such active mention of coherent artwork for explaining the artist's drawing style could give the user a more holistic view.

\subsubsection{Evaluation of LLM Response Correctness}
To assess the factual reliability of the LLM's conversational content, we conducted a post hoc manual analysis of the 529 visitor–AI interactions recorded during the experiment. The objective was to assess whether the model produced factually accurate statements about the exhibited artworks and to identify potential cases of misinformation or unverifiable descriptions.
Each response was first segmented into individual factual descriptions, referred to as marks. A mark represents a single factual unit in the model’s output—for example, statements such as “one of the Water Lilies series,” “painted at his home in Giverny,” or “a portrait of Monet’s wife Camille and their son.” All marks were independently annotated by three bilingual researchers with reference to art-historical literature and exhibition materials related to the ten Monet paintings included in the experiment. To ensure the reliability and transparency of the factual mark evaluation, we adopted a consensus-based qualitative annotation procedure. Prior to full-scale annotation, all annotators jointly reviewed a subset of LLM responses to calibrate the definition of a “factual mark” and to agree on annotation criteria, including the requirement that each response contain at least one objectively verifiable factual statement when applicable.

All annotations were conducted using Taguette \footnote{https://www.taguette.org/about.html}, an open-source collaborative qualitative analysis tool, which enabled traceable coding and systematic comparison across annotators. Three bilingual researchers independently annotated the segmented marks with reference to established art-historical sources. Disagreements were resolved through structured discussion until full consensus was reached, resulting in a final set of agreed annotations used for analysis.
Based on cross-referencing with verified sources, every mark was classified into one of three categories:

\begin{itemize}
\item \textbf{Correct marks} accurately reflected factual information regarding the artwork’s title, creation period, location, or depicted subject.
\item \textbf{Incorrect marks} contained factual inconsistencies, such as incorrect dates, places, or misidentification of the represented figures or objects.
\item \textbf{Unverifiable marks} referred to interpretive or affective statements that could not be confirmed or refuted through documentary evidence—for instance, comments about the artist’s emotions or the viewer’s subjective feelings.
\end{itemize}

Among the ten artworks, The Seine at Lavacourt was excluded from the analysis because its English title was mistakenly registered as The Seine near Giverny, leading the LLM to generate responses based on incorrect contextual information embedded in the initial prompt. The remaining nine paintings were retained for evaluation.
Table~\ref{tab:llm_correctness_by_artwork} summarizes the evaluation results for each artwork. 
As shown, the majority of the responses across paintings were judged to be factually accurate, 
while a smaller portion contained minor factual inconsistencies or unverifiable statements. 
The second artwork, "The Seine at Lavacourt", was excluded from the analysis due to an initial labeling error in the prompt, which caused all related responses to be based on incorrect contextual information. 
Rather than computing an overall accuracy rate across all paintings, 
the evaluation was conducted on a per-artwork basis to reflect the localized correctness of the LLM outputs more appropriately. 
Overall, the results suggest that the model maintained a generally high level of factual reliability in describing the artworks, 
with most inaccuracies stemming from occasional misidentifications of creation dates or depicted subjects. 
This evaluation provides a more transparent view of the model’s factual behavior and offers an empirical basis for refining prompt design and data grounding in future iterations of the system. It is important to note that this validation focuses on factual correctness at the mark level and does not aim to fully capture higher-level interpretive qualities, such as curatorial appropriateness, narrative coherence, or educational framing.

\begin{table}[H]
    \centering
    \begin{threeparttable}
    \caption{Post hoc evaluation of LLM response correctness by artwork. Each response was segmented into factual marks and annotated as Correct, Incorrect, or Unverifiable. Percentages are calculated within each artwork.}
    \small
    \renewcommand{\arraystretch}{1.2}
    \begin{tabular}{p{1.0cm} p{4.0cm} p{2cm} p{2.0cm} p{2.0cm} p{2.0cm}}
        \hline
        \textbf{No} & \textbf{Title of Painting} & \textbf{Total Marks} & \textbf{Correct (\%)} & \textbf{Incorrect (\%)} & \textbf{Unverifiable (\%)} \\ \hline
        1 & Poppy Field 
          & 89 
          & 53 (59.6) 
          & 24 (27.0) 
          & 12 (13.5) \\ \hline
        2 & The Seine at Lavacourt 
          & 133 
          & -- 
          & -- 
          & -- \\ \hline
        3 & The Water Lily Pond 
          & 134 
          & 128 (95.5) 
          & 3 (2.2) 
          & 3 (2.2) \\ \hline
        4 & The Magpie 
          & 135 
          & 115 (85.2) 
          & 16 (11.9) 
          & 4 (3.0) \\ \hline
        5 & Water Lilies Weeping Willows 
          & 212 
          & 211 (99.5) 
          & 1 (0.5) 
          & 0 (0.0) \\ \hline
        6 & Sunset on the Seine 
          & 202 
          & 173 (85.6) 
          & 25 (12.4) 
          & 4 (2.0) \\ \hline
        7 & Woman with a Parasol 
          & 247 
          & 219 (88.7) 
          & 25 (10.1) 
          & 3 (1.2) \\ \hline
        8 & Cliff Walk at Pourville 
          & 114 
          & 101 (88.6) 
          & 7 (6.1) 
          & 6 (5.3) \\ \hline
        9 & Impression, Sunrise 
          & 91 
          & 86 (94.5) 
          & 5 (5.5) 
          & 0 (0.0) \\ \hline
        10 & Water Lilies 
          & 83 
          & 75 (90.4) 
          & 3 (3.6) 
          & 5 (6.0) \\ \hline
    \end{tabular}
        \footnotesize\textit{Note: - indicates that data corresponding to Artwork No.~2 were not included in this analysis due to an initial labeling mismatch between two visually similar artworks (The Seine at Lavacourt and The Seine near Giverny), which limited the interpretability of LLM-based results.
}
    \label{tab:llm_correctness_by_artwork}
    \end{threeparttable}
\end{table}

\subsection{Semi-Structured Interview Findings}

To synthesize participant reflections from the 26 semi-structured interviews, we employed a bottom-up, inductive thematic coding process. Through iterative coding and collaborative review, we identified a series of recurring concepts, which were subsequently grouped into three overarching thematic dimensions:
(1) \textbf{Overall (WhiteTesseract System) Evaluation and Future Vision},  
(2) \textbf{LLM Conversational Interaction Experience}, and  
(3) \textbf{XR Visual Immersion Experience}.

\begin{figure}[ht]
    \centering
    \includegraphics[width=1.0\linewidth]{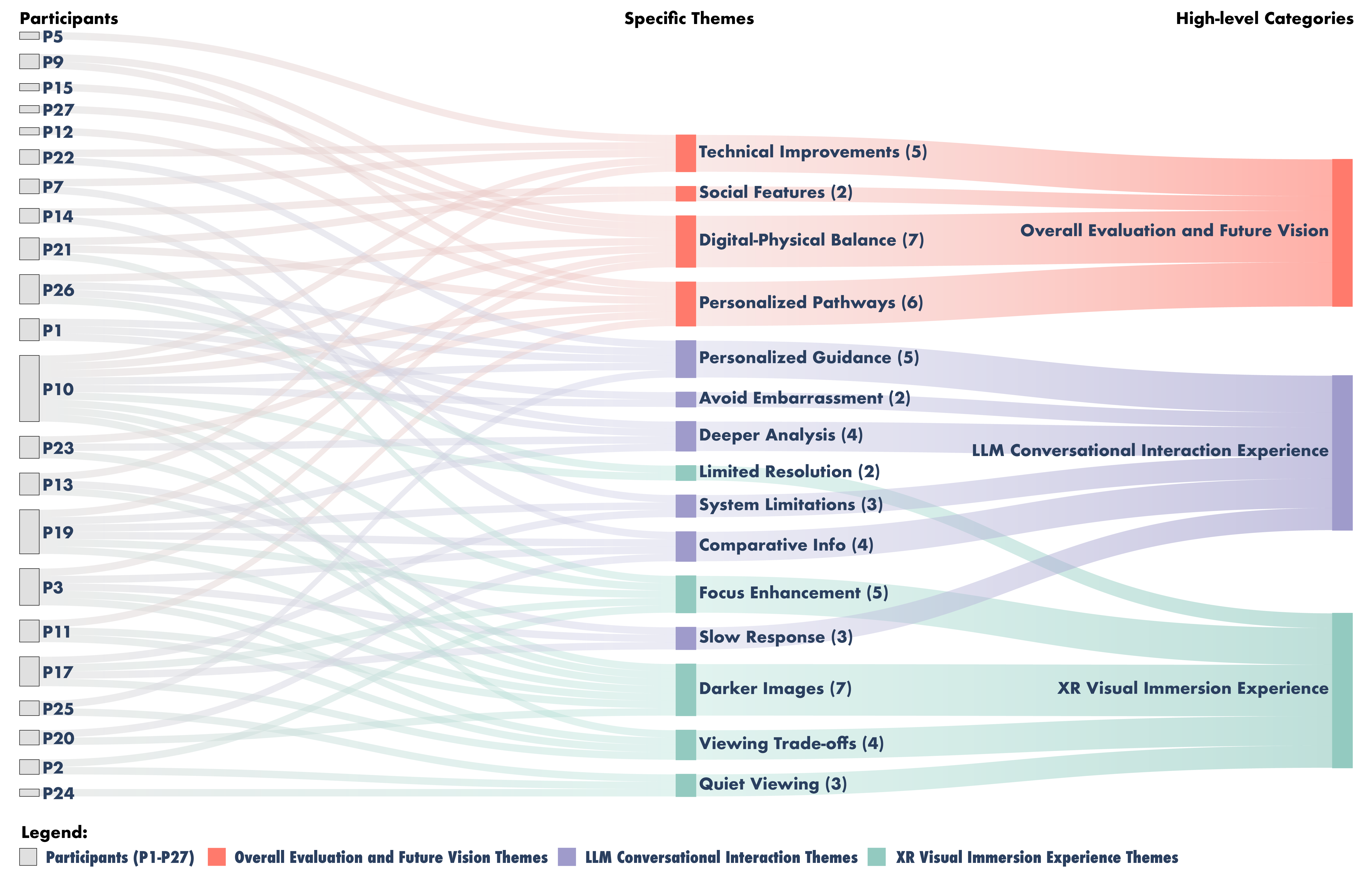}
    \caption{Semi-structured interview themes: from participant responses to thematic categories. Note: Darker shades indicate high-level categories.}
    \label{fig:Thematic}
\end{figure}

To facilitate cross-participant comparison and illustrate the structural relationships between specific codes and higher-level categories, we created a Sankey diagram (Fig. \ref{fig:Thematic}). This visualization maps the flow of responses from individual participants (left) to specific subthemes (center) and ultimately to aggregated thematic categories (right). The numbers in parentheses represent the number of participants who contributed to each subtheme. Fig. \ref{fig:ques_response} reveals the diverse and often overlapping ways in which participants engaged with the system. While some reflected primarily on sensory or perceptual factors, others focused on the responsiveness and depth of the conversational AI or speculated about future forms of interactive art engagement. Notably, most participants contributed to multiple thematic areas, highlighting the interconnectedness of \textit{immersion}, \textit{interpretive agency}, and \textit{technological usability}.

Participants reflected extensively on how the XR system shaped their perceptual engagement with artworks. While several noted visual limitations—such as darker image rendering, color distortion (P3, P10, P11, P13, P19, P20, P23), and insufficient resolution to capture fine detail (P10, P26), many appreciated the system’s ability to reduce distractions and support sustained focus on individual pieces. The “diminished reality” effect was frequently described as helpful in facilitating deeper attention and interpretive clarity (P2, P10, P17, P19, P21), with some likening it to a quiet, undisturbed viewing experience that differed from traditional gallery settings (P2, P24, P25). At the same time, several participants expressed concerns about the trade-offs of this immersive design. Unlike conventional exhibitions that allow side-by-side viewing and comparison, the system’s isolated focus on single works limited peripheral awareness (P3, P10) and, for some, created a sense of spatial disconnection or orientation loss (P10, P11, P17).

In parallel, participants evaluated the conversational AI component in terms of personalization, responsiveness, and its potential for meaningful interaction. Many respondents praised its ability to deliver tailored guidance and more engaging dialogue than conventional audio guides (P1, P10, P22, P25, P26), with some highlighting how the virtual interface helped reduce the social pressure associated with asking questions aloud (P1, P10). Nevertheless, several participants encountered issues with response latency, repeated answers, and limited comprehension of nuanced queries (P3, P7, P13, P17, P19), which at times disrupted the fluidity of interaction and reduced their trust in the system’s intelligence. These shortcomings did not dampen broader expectations: multiple participants envisioned future iterations offering deeper analytical content (P1, P19, P23, P26), including comparative insights across artworks or artists (P3, P14, P19, P20). Others imagined more socially dynamic forms of interaction, such as art dialogue modeled after personal interests or collaborative conversation with peers (P10, P22).

\subsection{Quantitative Results}

\subsubsection{Impact on exhibition viewing behavior}
\begin{figure}[ht]
    \centering
    \includegraphics[width=1\linewidth]{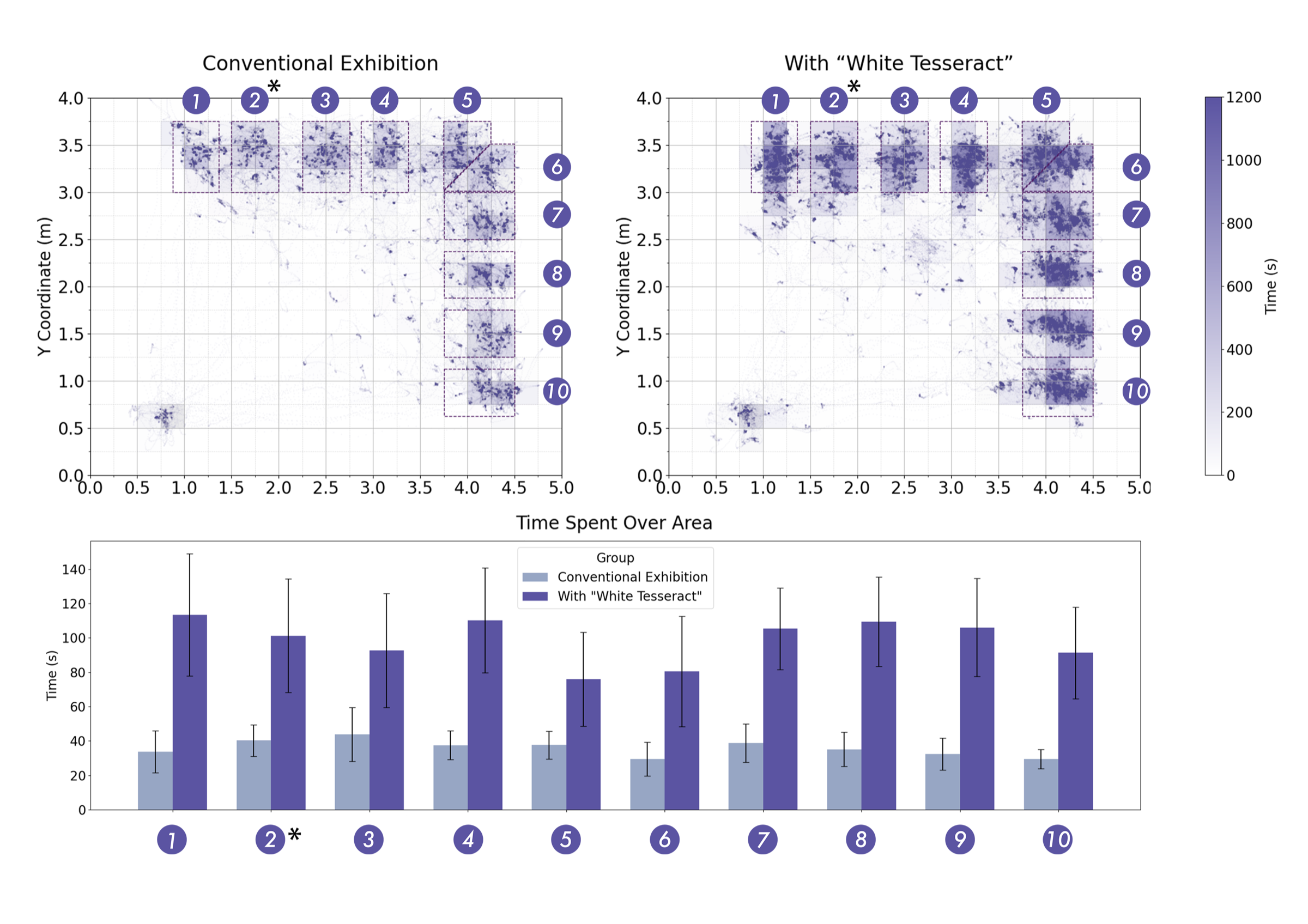}
    \caption{Motion tracking data analysis visualization of two conditions. The areas' partition refers to the position of each paintings. (a) shows the cumulative position distributions in traditional visiting and visiting with our system. (b) is the statistic results of the time spending over each area.}
    \raggedright
\footnotesize\textit{Note: * indicates an initial title mismatch for Artwork No.~2 (The Seine at Lavacourt and The Seine near Giverny).}
    \label{fig:Time-domainA}
\end{figure}

Movement data was recorded using the HTC Vive Tracker (first generation) and its base stations to capture the participants' positions accurately. The positions of artworks served as spatial references to define the viewing areas (1 to 10). Movement trajectories were projected onto a top-down view to calculate the time spent viewing each artwork. To determine the total time spent in each designated area coordinates with corresponding timestamps were analyzed by identifying entry and exit points within predefined regions outlined by polygonal boundaries. Since participants could revisit the same artwork multiple times, the time spent during each visit was cumulatively summed to obtain the total viewing duration for each area (Fig. \ref{fig:Time-domainA} a).

Previous studies provided a baseline on the average time individuals spend viewing artwork. We compared the average time spent of the control group with previous studies, and a similar result was observed: Our study: M = 35.3s, \( \text{Mdn} \)= 35.2s; Smith-Smith's study: M = 27.2s, \( \text{Mdn} \)= 17.0s; \cite{Smith2001}, Carbon's study: M = 50.5s, \( \text{Mdn} \)= 43.0s\cite{Carbon2017}. This alignment validates the exhibition settings. Results of the time domain analysis are shown in Table \ref{tab:statistical_significance}.

\begin{table}[ht]
\centering
\caption{Statistical Significance of Viewing Duration Across Areas}
\begin{tabularx}{\textwidth}{l>{\centering}X>{\centering}X>{\centering}X>{\centering\arraybackslash}X}
\toprule
\textbf{Area} & 
\shortstack{Mean \\ (Conventional) (s)} & 
\shortstack{Mean \\ ("Tesseract") (s)} & 
\shortstack{p-value} & 
\shortstack{95\% CI \\ (Group 1 / Group 2)} \\
\midrule
NO. 1  & 33.7  & 113.3  & \( <0.001 \) & (21.5, 45.9) / (77.7, 148.8) \\
NO. 2  & -  & -  & - & (-, -) / (-, -) \\
NO. 3  & 43.8  & 92.6  & 0.030 & (28.2, 59.5) / (59.5, 125.7) \\
NO. 4  & 37.5  & 110.2  & \( <0.001 \) & (29.2, 45.8) / (79.7, 140.8) \\
NO. 5  & 37.6  & 75.9  & 0.048 & (29.5, 45.8) / (48.5, 103.3) \\
NO. 6  & 29.5  & 80.5  & 0.034 & (19.8, 39.2) / (48.3, 112.7) \\
NO. 7  & 38.7 & 105.3  & \( <0.001 \) & (27.6, 49.9) / (81.6, 129.1) \\
NO. 8  & 35.2  & 109.4  & \( <0.001 \) & (25.3, 45.1) / (83.4, 135.5) \\
NO. 9  & 32.4  & 106.0  & 0.001 & (23.0, 41.7) / (77.5, 134.5) \\
NO. 10 & 29.4  & 91.2  & 0.001 & (23.9, 35.0) / (64.5, 118.0) \\
\bottomrule
\end{tabularx}
\raggedright
\footnotesize\textit{
Note: - indicates that viewing duration data for Area No.~2 were not included in the statistical analysis due to an initial labeling mismatch between two visually similar artworks (The Seine at Lavacourt and The Seine near Giverny).
}
\label{tab:statistical_significance}
\end{table}

Results revealed significant differences in viewing time between the conventional way and the "WhiteTesseract" experience across all areas, as detailed in Table \ref{tab:statistical_significance}. The mean viewing times under conventional conditions ranged from 29.44 seconds (Area 10) to 43.84 seconds (Area 3), comparing the experimental group, which ranged from 75.92 seconds (Area 5) to 113.28 seconds (Area 1). Statistical analysis using the Mann-Whitney U test indicated significant differences between the two conditions. Highly significant results were observed in Areas 1, 4, 7, and 8 (\textit{p}<0.001), while other areas also showed statistical differences, with p-values ranging from 0.001 to 0.048. The 95\% confidence intervals further support these findings. 

\subsubsection{System and Experience Evaluation}
\begin{figure}[H]
    \centering
    \includegraphics[width=1.0\linewidth]{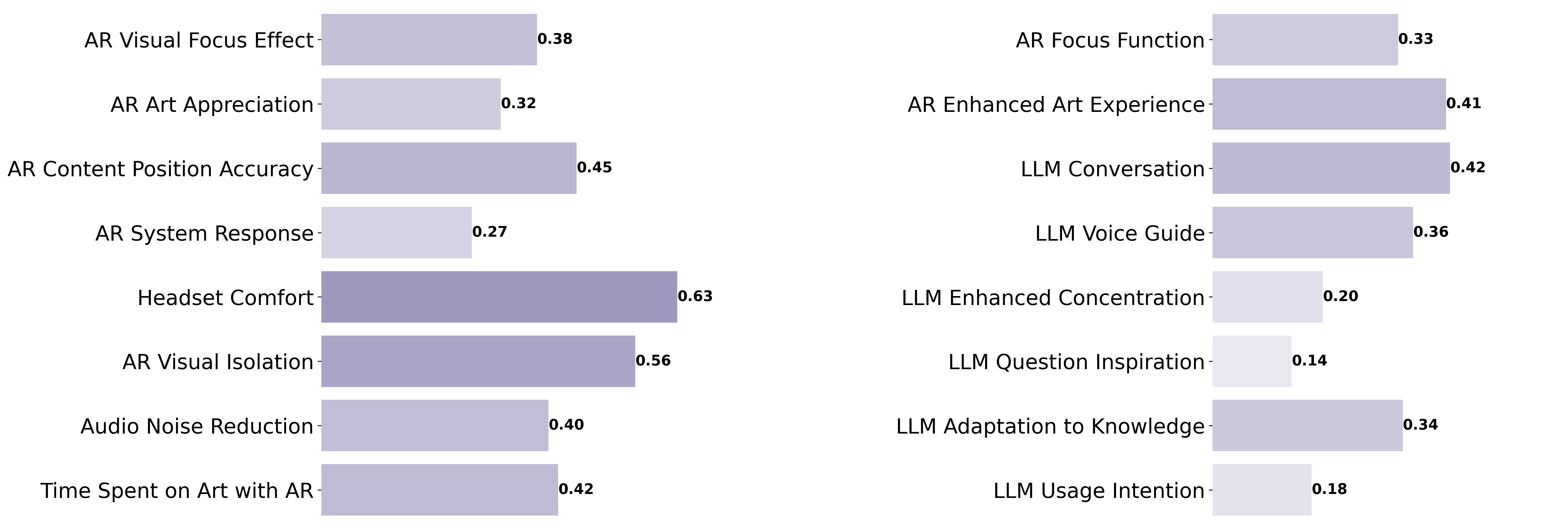}
    \caption{Spearman Correlations between System Features and UEQ Total Scores}
    \label{fig:spearman}
\end{figure}

The SUS data shows a mean score of 29.70 (total 50), with a standard deviation of 4.11. The 95\% confidence interval ranges from 28.08 to 31.33. This result reveals that the usability of the system was not ideal. Potential reason like "\textit{The prolonged use of the heavy headset leads to fatigue and neck discomfort.}" has been mentioned a lot in the final interviews. 
The average total score of UEQ is 30.30 with a standard deviation of 5.18, the 95\% confidence interval is (28.25, 32.35).  To further understand what influence factors are important for better user experience, we did a Spearman's rank correlation coefficient between the UEQ total score and customized questionnaires. As shown in Fig. \ref{fig:spearman}, the analysis reveals varying degrees of association between different questions and the UEQ total score - higher correlation coefficients indicate a stronger monotonic relationship. 

\section{DISCUSSION}

Our analysis integrates quantitative behavioral data with qualitative insights from semi-structured interviews conducted with all 26 participants immediately following their experimental sessions. These interviews, lasting 15-20 minutes each, employed open-ended questions to elicit participants' subjective experiences, interpretive processes, and reflections on both the perceptual modulation and AI dialogue components of WhiteTesseract. The interview protocol progressed from general impressions to specific probes about attention management, question formulation strategies, and perceived limitations of the system. This mixed-methods approach enabled us to move beyond surface-level behavioral metrics to understand the cognitive and affective mechanisms underlying the observed changes in viewing patterns and interpretive engagement. By triangulating motion tracking data, conversation logs, and participant narratives, we can examine what changes occurred in museum viewing behavior, why participants adopted particular strategies, and how they made sense of the novel exhibitionary format. The following sections synthesize these multiple data streams to address our research questions concerning the WhiteTesseract system’s overall behavioral impact (RQ1) and the role of AI-facilitated dialogue in shaping interpretive engagement (RQ2).

\subsection{RQ1: Impact of WhiteTesseract System on Viewing Behavior}

\subsubsection{Extended Viewing Duration through Perceptual Modulation}
The observed behavioral differences reflect the overall influence of the WhiteTesseract experience, where visual attention modulation and AI dialogue jointly shape visitors’ engagement dynamics. Our findings demonstrate that WhiteTesseract design fundamentally transformed participants' viewing behavior, most notably through a dramatic extension of viewing duration. The quantitative data revealed that mean viewing time increased from 35.3 seconds in the conventional condition to 98.3 seconds with WhiteTesseract across all artworks, representing a 175\% increase. This effect was consistent across individual artworks, with some areas showing even more pronounced differences, Area 1, for instance, increased from 33.72s to 113.28s (\textit{p} < 0.001).

This substantial increase in viewing duration can be attributed to the system's ability to create what participants repeatedly described as a "personal attention bubble." Through visual masking and spatial sound isolation, visitors experienced reduced environmental distractions, while conversational prompts encouraged sustained interpretive reflection. As P3 noted, "Everything else turned white, so I felt gently compelled to concentrate on just one piece." Several participants (P1, P15, P17, P18) also mentioned that being able to “talk” with the system deepened their focus by giving them reasons to linger and think, turning passive looking into active interpretation.

The mechanism underlying this extended engagement aligns with established theories of attention design in HCI and cognitive psychology. By reducing external cognitive load through environmental modulation, the system enabled what Kahneman would term "effortless attention", a state in which cognitive resources are not depleted by filtering distractions but can be fully devoted to interpretive processing \cite{kahneman1973attention, wickens2022applied}. These findings extend prior results showing that when perceptual and conversational elements co-exist, visitors sustain both perceptual focus and reflective reasoning longer than in static environments \cite{brieber2015white}. 

\subsubsection{Changes in Movement Patterns and Exhibition Navigation}
Beyond temporal extension, WhiteTesseract altered the spatial dynamics of exhibition navigation. The motion tracking data revealed that participants adopted fundamentally different movement patterns when using the system. In conventional viewing, participants tended to follow a relatively linear path through the gallery, spending comparable time at each artwork (ranging from 29.44s to 43.84s). With WhiteTesseract, however, viewing became more selective and deliberate, with some artworks receiving substantially more attention than others (ranging from 75.92s to 113.28s).

This shift from uniform to selective engagement suggests that the system allowed visitors to construct personalized pathways through the exhibition based on their individual interests and curiosity. Several participants explicitly noted this change in their interview responses. P23 described it as "choosing the pace and the depth—it felt more like my own exhibition." The ability to isolate individual artworks perceptually appeared to liberate visitors from the implicit social pressure to maintain steady progression through the gallery, allowing them to dwell with works that resonated more deeply \cite{falk2016museum}.

However, this personalization came with trade-offs. Some participants noted that the immersive mode made it challenging to maintain an overall awareness of the exhibition. P10 and P11 mentioned that "in the immersive mode, it would be difficult to see multiple artworks simultaneously," suggesting that while WhiteTesseract enhanced focused engagement with individual works, it potentially fragmented the holistic experience of the exhibition as a curated ensemble. This tension between depth and breadth represents a fundamental design challenge for future adaptive exhibition systems.

\subsection{RQ2: AI-Dialogue's Effect on Interpretive Engagement}

\subsubsection{Diversification of Question Types and Interpretive Autonomy}

While participants appreciated the dialogic nature of the interaction, our findings reveal a complex relationship between perceived depth and actual interpretive substance. Analysis of the 529 total questions revealed a diverse distribution that markedly differs from the information typically provided by traditional museum labels. Where conventional wall texts primarily offer descriptive information (artwork title, date, medium, dimensions) and basic contextual details (historical period, artistic movement), our participants' questions showed a much broader interpretive scope \cite{leinhardt2004listening}: Descriptive Information (40.8\%), Analytical Understanding (19.61\%), Emotional and Personal Engagement (9.84\%), Comparative and Contextual Inquiry (12.69\%), Technical and Functional Questions (6.7\%), and Open-Ended Exploration (10.36\%).

This distribution is particularly striking when we consider that traditional labels rarely address emotional responses, personal connections, or comparative analysis across works \cite{serrell2015exhibit}. The fact that nearly 60\% of the questions went beyond basic descriptive information—encompassing analytical, emotional, comparative, and exploratory dimensions-suggests that the AI dialogue system enabled visitors to pursue interpretive paths that static labels cannot support. For instance, questions like "Did the model of the child, his wife Camille, and son Johnny have to stand there the whole time while painting?" or "Why was Monet fascinated by Japanese culture?" represent the kind of personal curiosity and cultural inquiry that traditional labels, constrained by space and universal relevance, typically omit. Many participants reported that this expanded scope encouraged them to think more deeply about the artworks—P5 noted that "AR made me focus more on a single piece and think about its meaning," while P19 and P23 specifically mentioned that the LLM could provide "deeper analysis of the questions." The ability to ask follow-up questions and receive immediate responses created what P26 described as a more "interactive" experience that fundamentally transcended the one-way information delivery of traditional interpretation.

The AI was widely perceived not as a teacher or expert but as a "gentle companion"—someone to talk to without fear of making mistakes. This perception was deeply rooted in the system's non-judgmental tone and anonymous interaction model. As P1 noted, "It felt like I could ask anything without being embarrassed, even questions I would never ask a human guide." Participants such as P4 and P11 emphasized the system's ability to support dialogue "with anyone at any time," reinforcing that conversational openness and emotional safety lowered the threshold for engagement. This finding aligns with prior research on the value of non-evaluative social agents in educational contexts, where agents designed as learning companions rather than authoritative tutors have been shown to increase willingness to ask questions and foster continued engagement \cite{kim2006social, kim2006pedagogical}.

\subsubsection{Perceived Depth versus System Limitations}

However, this comfort was not always paired with cognitive depth. The word frequency analysis of LLM responses revealed repetitive use of terms like "light" and "color," suggesting that while the system maintained conversational fluency, it often defaulted to generic artistic descriptors rather than providing genuinely insightful analysis. Many participants found the answers to be repetitive, vague, or disconnected from the artwork itself. As P5 reflected, "It responded, but not in a way that gave me more to think about." Others expressed disappointment when the system failed to recognize the specificity of their questions, P13 and P14 noted frustration when comparisons across artworks fell short, while P19 articulated, "I wanted the AI to go deeper... to answer from multiple angles, not just repeat what's on the label." In these cases, the AI's role as a "companion" began to feel superficial, warm in tone but lacking substance.

This gap between conversational presence and interpretive substance was not simply a matter of interaction design but a limitation rooted in the LLM's underlying capabilities \cite{bender2020climbing}. Several participants (P5, P13, P14) noted that the AI lacked the knowledge specificity needed to ground responses in historical, stylistic, or comparative art contexts. The system struggled particularly with the 12.69\% of questions seeking comparative and contextual inquiry, often providing generic responses when participants sought nuanced connections between works or artistic movements. This reflects a structural challenge in using general-purpose LLMs in domain-specific interpretive settings: rather than lacking specialized art historical or curatorial knowledge, these models tend to be "lazy," often not actively retrieving relevant details unless explicitly prompted. Such passivity limits their effectiveness in museum contexts, where divergent and proactive exploration of information can significantly enhance visitors' interpretive engagement with artworks.

Technical disruptions further exacerbated this tension between accessibility and depth. Participants P17 and P18 reported voice recognition issues, delays, or misinterpretations that interrupted the conversational flow and diminished the system's perceived intelligence. P3 and P13 specifically mentioned that "the reaction of the system was sometimes slow and difficult to interrupt," while P7, P17, and P19 noted that "the system could not accurately understand questions or give similar answers." These moments highlighted a critical fragility: trust in dialogic systems can quickly erode when interactional continuity or semantic clarity breaks down \cite{luger2016like}. The technical limitations were particularly problematic for the 6.7\% of questions that were technical or functional in nature, where users expected precise system responses but often encountered confusion or delays.

Beyond stylistic repetition, a subtler limitation emerged in the factual grounding of the LLM’s responses. Our post hoc verification (Section 6.1.3) revealed that, while most statements were accurate, occasional hallucinations occurred—such as misattributing a painting’s location or interpreting artistic intent as factual detail. To minimize such drift, each artwork was pre-labeled with its title and metadata before being sent to the LLM, effectively constraining the model’s interpretive space and reducing spontaneous hallucinations common in open-ended dialogue. However, this mechanism also created a trade-off between control and adaptability: once the input metadata was incorrect, as in the case of The Seine at Lavacourt, mistakenly labeled as The Seine near Giverny, the entire chain of responses became misdirected and the model could not self-correct. This design thus improved factual reliability under accurate grounding but amplified systemic errors when contextual information was flawed, highlighting the need for future iterations to integrate dynamic verification methods and domain-specific datasets capable of detecting and recovering from such inconsistencies in real time.

Beyond system-level constraints related to factual grounding and error propagation, the current study also has limitations at the level of evaluation scope. While an overall qualitative judgment by a domain expert (e.g., curators or art historians) could provide valuable insight into interpretive quality, such assessments were beyond the scope of the current study, which focuses on system feasibility and interaction patterns. We therefore adopt a conservative validation strategy centered on factual correctness, treating it as a necessary but not sufficient condition for interpretive quality. Future work will be required to incorporate expert-led qualitative evaluations to assess narrative appropriateness, contextual sensitivity, and educational value in museum-specific settings.

Despite these limitations, the distribution of question types and participant feedback suggests that the AI dialogue system succeeded in broadening the scope of visitor engagement beyond what traditional labels typically support. The presence of emotional engagement questions (9.84\%) and open-ended exploration (10.36\%) indicates that participants felt empowered to pursue personal and creative lines of inquiry. As P10 and P22 suggested in their vision for future systems, there is potential for AI companions that can adapt their response depth and style based on visitor expertise and interest, moving beyond current limitations to provide truly personalized interpretive experiences. The challenge for future systems lies in balancing the approachability and non-judgmental interaction style that made WhiteTesseract's AI perceived as a "gentle companion," with the depth of knowledge and responsiveness that can sustain meaningful intellectual engagement throughout the museum visit.

\subsection{Synthesis: Implications for Adaptive Exhibition Design}

\subsubsection{Toward Configurable Exhibition Spaces}

\begin{figure}[ht]
    \centering
    \includegraphics[width=1.0\linewidth]{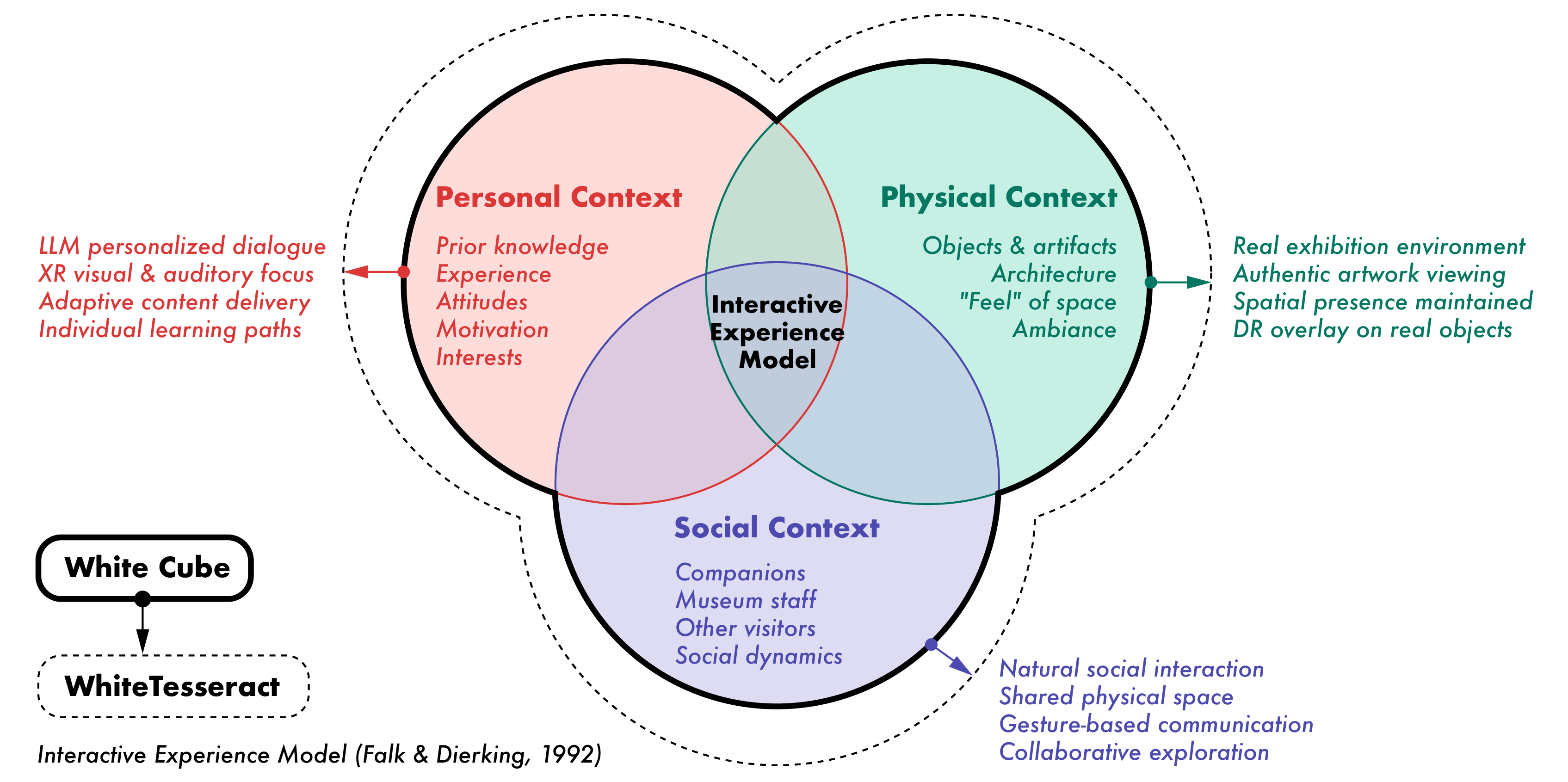}
    \caption{WhiteTesseract within the Interactive Experience Model Framework (adapted from Falk \& Dierking, 1992).}
    \label{fig:Framework}
\end{figure}

Our findings align with Falk and Dierking’s Interactive Experience Model (1992) \cite{dierking1992redefining}, which conceptualizes museum experiences as the intersection of three interdependent contexts: personal, physical, and social. This model provides an analytical framework for examining how emerging technologies influence visitor interactions within exhibition contexts, particularly by integrating personal, spatial, and social dimensions of experience. As illustrated in Fig~\ref{fig:Framework}, WhiteTesseract operates across all three domains. It enhances the personal context by offering adaptive AI-mediated dialogue and individualized perceptual control, allowing visitors to construct meaning in alignment with their prior knowledge, interests, and emotional states. It transforms the physical context through diminished reality techniques that modulate spatial perception and visual salience, enabling deeper engagement with individual artworks while minimizing environmental distractions. It also subtly reshapes the social context by supporting private, low-pressure interactions—particularly valuable for visitors who may feel self-conscious or inhibited in group settings—while preserving opportunities for shared reflection and discussion. By actively engaging with each of these experiential layers, WhiteTesseract demonstrates how adaptive exhibition systems can support more inclusive, exploratory, and personalized forms of cultural participation.

The convergence of our findings across both research questions reveals fundamental insights about the future of exhibition design in the era of high-resolution XR. WhiteTesseract demonstrates that the integration of perceptual modulation (RQ1) and AI-mediated dialogue (RQ2) can create exhibition experiences that are simultaneously more focused and more expansive than traditional formats—more focused in terms of sustained attention to individual works, yet more expansive in the range of interpretive engagement they support.
The quantitative transformation in viewing behavior is striking: mean viewing duration increased from 35.8 seconds to 98.6 seconds, with consistent effects across all artworks. However, this temporal extension represents only the surface of a deeper phenomenological shift. When visitors can control their sensory environment through diminished reality and engage in open-ended dialogue about artworks, they transition from passive viewers following predetermined paths to active constructors of their own exhibition experience \cite{dourish2001action}. As P23 articulated, it became "more like my own exhibition," suggesting that adaptive technologies can fundamentally alter the power dynamics between visitors, artworks, and curatorial authority. This shift toward personalization was echoed by P9, who noted that while current technology offers just ordinary viewing, technological advances in the coming years will make these experiences truly transformative and wearable.

The diversification of interpretive engagement further reinforces this shift toward visitor agency. Where traditional museum labels constrain interpretation to basic descriptive and contextual information, our participants' 529 questions spanned six distinct categories, with nearly 60\% moving beyond factual information to explore analytical, emotional, comparative, and speculative dimensions. This finding suggests that when given appropriate tools, visitors naturally seek deeper, more personalized forms of meaning-making than static interpretation allows. The AI's role as a "gentle companion" lowered psychological barriers to inquiry, enabling visitors to ask questions that they might never voice in traditional settings. As P15 observed when comparing exhibition formats, visitors actively contemplated the balance between digital enhancement and physical authenticity, questioning which approach best serves the art experience.

Yet our findings also illuminate critical tensions that must guide future development. The trade-off between focused attention and holistic awareness presents a fundamental design challenge: while perceptual isolation enhanced individual artwork engagement, it risked fragmenting the exhibition's curatorial narrative. P19 explicitly addressed this concern, expressing desire to maintain connection with the real physical space even when using mixed reality. P26 envisioned a hybrid approach where visitors might preview exhibitions digitally at home before experiencing the physical artworks, suggesting that digital and physical encounters could be complementary rather than competitive. Similarly, while the AI dialogue system successfully broadened interpretive scope, its limitations in providing substantive, contextually grounded responses revealed the gap between conversational fluency and genuine expertise.

Moving forward, we propose that adaptive exhibition systems should be conceived not as replacements for traditional formats but as configurable layers that visitors can modulate based on their needs and preferences. This vision aligns with participants' own imaginings of future possibilities. P11 envisioned an AI companion that could engage in collaborative thinking about how to deepen the understanding of artworks, while P12 imagined systems where personal preferences could actively shape the creative experience. Future systems might offer multiple "attention modes"—from full perceptual isolation for deep contemplation to augmented awareness that maintains peripheral vision while highlighting curatorial connections. Similarly, AI interpretation could adapt dynamically, offering different depths of engagement ranging from basic orientation for first-time visitors to scholarly discourse for experts.

\subsubsection{Balancing Individual Agency and Collective Culture}

The implications extend beyond individual system design to institutional practices. Museums adopting such technologies must reconsider fundamental assumptions about curatorial authority, visitor agency, and the social dynamics of gallery spaces \cite{vom2001exhibiting}. If every visitor can create their own perceptual bubble and engage in personalized dialogue, what becomes of the shared, collective experience that museums traditionally foster? Several participants grappled with this question. P20 expressed desire to discover related works and understand shared techniques across the exhibition space, suggesting that personalization should enhance rather than replace curatorial connections. P17 similarly emphasized wanting easy navigation to find other works, indicating that individual focus should not come at the cost of serendipitous discovery.
Our findings suggest that the answer lies not in choosing between individual and collective experience but in designing systems that support fluid transitions between different modes of engagement. P7 captured this balance elegantly, noting that while pure digital experiences might excel for educational purposes, mixed physical-digital exhibitions have broader public appeal. P13 articulated a vision where visitors could move fluidly between analytical views (diagrams and technical details) and emotional engagement with individual paintings and their natural contexts. WhiteTesseract points toward a future where visitors might begin in social, exploratory mode, shift to isolated contemplation for works that particularly resonate, engage in AI-mediated inquiry to deepen understanding, and then return to shared discussion enriched by their personalized exploration.

The social dimension of future exhibitions also emerged as a key consideration. P14 imagined exhibitions where visitors could see virtual representations of other viewers and even engage in shared interpretation, while P21 envisioned leaving digital traces or messages for future visitors. These suggestions point toward hybrid spaces that maintain social presence while enabling individual customization. As P3 emphasized, the future lies not in a single model but in supporting diverse choices and approaches to experiencing art.

Ultimately, WhiteTesseract demonstrates that high-resolution XR and AI can do more than augment traditional exhibitions; they can reimagine the fundamental relationship between visitors and artworks. By giving visitors control over both perceptual conditions and interpretive dialogue, these technologies support a more active, exploratory, and personally meaningful form of cultural engagement. The challenge for the museum field, as P27 astutely observed, is ensuring that technology supports rather than replaces authentic encounters with art. As we continue to develop adaptive exhibition systems, the goal should not be to maximize individual customization at all costs but to create flexible frameworks that honor both personal agency and collective culture. The future of exhibitions lies not in choosing between physical and digital, individual and social, or focused and exploratory modes, but in creating adaptive systems that allow visitors to move fluidly between these states based on their evolving needs and interests.

\subsubsection{Practical Considerations for Real-World Deployment}
While our controlled study demonstrates the system’s conceptual effectiveness, its translation into real museum settings raises a number of operational and logistical challenges. Each Apple Vision Pro costs approximately \$3,500, making scalable deployment prohibitively expensive for most institutions. Device management introduces operational challenges: our study required 5-10 minutes per participant for calibration, which would create bottlenecks during peak hours. Battery life (2.5-3 hours) limits extended visits, and hygiene protocols for shared headsets further reduce throughput. Beyond hardware, deployment also involves software-related expenses, including maintaining the local server environment, ensuring compatibility with evolving XR platforms, and refining dialogue behavior through iterative testing. As a result, large public museums may face substantial financial and staffing barriers to adoption, whereas smaller galleries, university museums, and temporary exhibitions could more feasibly pilot such systems under controlled conditions. Hybrid configurations, such as limited “immersive viewing sessions” or optional booking-based experiences, may represent a practical middle ground between innovation and scalability.

Furthermore, while the current system was implemented as a single-user experience, it still allows visitors to converse and interact freely with their companions in the shared physical space. However, others cannot join or access the same AI-mediated dialogue environment. Many museum visitors attend exhibitions in small groups, often with family members, friends, or partners, and their interactions, including shared gaze, gesture, and conversation, play a crucial role in meaning-making and collective engagement \cite{Falk2016Motivations, christidou2010re, COVES2025}. This design choice, therefore, prioritizes focused personal reflection within the system but limits opportunities for shared interpretive exploration.

Our 5m × 4.5m controlled space with sequential sessions differs fundamentally from major institutions where hundreds of visitors simultaneously navigate interconnected galleries. WhiteTesseract's requirement for spatial tracking and one-to-one device allocation makes it better suited for temporary exhibitions with timed entry (10-15 visitors per session), small galleries, or private educational programs rather than large-scale public museums.

The most pragmatic path forward may position WhiteTesseract as optional enhancement rather than default experience. Galleries could offer designated "immersive viewing sessions" at specific times, preserving collective engagement while enabling deeper individual contemplation. As P7 noted, mixed physical-digital formats demonstrate broader appeal than purely digital experiences. Future development should investigate lightweight AR glasses to reduce cost and discomfort, asynchronous social features that maintain focus while enabling delayed interaction, and flexible attention modes allowing rapid switching between isolated and social viewing. These implementation constraints concern scalability and operational logistics, whereas broader methodological and contextual limitations are discussed in the following section.

\subsection{Limitations and Generalizability}
Beyond these practical and logistical barriers, our controlled experimental design also introduced methodological constraints that limit the generalizability of the findings. The gallery environment was deliberately simplified: consistent lighting, minimal architecture, reproductions rather than originals, and absence of other visitors. Real museums present complex sensory environments where natural light varies, architectural features shape movement, and the material presence of authentic works generates engagement that reproductions cannot replicate \cite{bitgood2016attention}. Our finding that viewing duration increased from 35.8s to 98.6s occurred within this simplified context. Although the study provides insights into the integrated experience of WhiteTesseract, the current within-subjects design does not disentangle the specific contributions of its visual and dialogic components. Future work will isolate these effects to further validate the attention mechanism’s role. Moreover, because participants could freely adjust their level of perceptual isolation, minor variability across sessions may have introduced uncontrolled differences in attentional focus. Future studies should employ a standardized isolation setting to reduce potential confounding variables and strengthen the reliability of the findings.

The single-visitor protocol eliminated social dynamics that profoundly influence museum behavior. Participants experienced the gallery alone, free from companions or crowding. Research shows that social cues, such as companion presence, observations of others, and implicit pacing norms, significantly shape individual behavior \cite{Falk2016Motivations, christidou2010re, COVES2025}. The extended viewing times may partly reflect absence of these regulatory factors rather than solely system effects. How WhiteTesseract performs when visitors must negotiate shared space and companions' interests remains unknown. 

Our participant sample (N=27, ages 20-30, 92\% engineering backgrounds, 100\% prior XR experience) represents early adopters rather than general museum audiences. This early adopter profile may have influenced multiple outcomes reported in this study. First, participants with prior XR exposure may show lower initial cognitive load during onboarding and greater tolerance to headset-related constraints, leading to longer voluntary engagement and extended dwell time compared to first-time XR users \cite{o2012understanding}. Second, a technically literate population may be more inclined to probe system behavior, ask “how/why” questions, and iterate on prompts, which could shift the observed distribution of question types and increase the apparent diversity of dialogue \cite{gunatilake2024enablers}. Third, prior familiarity with interactive AI may shape participants’ expectations and evaluation criteria, potentially affecting subjective judgments of usefulness and interpretive support \cite{lindblom2025take}. We therefore interpret the observed effects primarily as evidence of feasibility and interaction patterns in an early adopter setting; verifying the magnitude and form of these effects in heterogeneous museum audiences remains an important next step. Older adults, children, and XR unfamiliar visitors may experience higher cognitive load, longer adaptation, or different comfort levels \cite{ahmed2025human, weech2019presence}. Additionally, our choice of familiar Monet reproductions may not generalize to contemporary art, non-Western heritage, or less canonical works where baseline engagement differs. Some observed mechanisms may nevertheless be less sensitive to technical expertise, for example, the benefit of a controlled sensory environment for reducing external distractions, and the role of private dialogue in lowering social barriers to asking questions in gallery. In contrast, measures closely tied to exploratory behavior (e.g., voluntary dialogue frequency, prompt iteration, and the depth of analytical questioning) are more likely to vary with prior XR/AI familiarity \cite{radianti2020systematic}. We therefore caution that the relative ranking of effects may generalize more readily than their absolute magnitudes.

We captured only short-term effects within 75-100 minute sessions, with no assessment of retention or long-term impact. Whether WhiteTesseract's effects persist beyond immediate experience or influence subsequent visits cannot be determined. Technical constraints also shaped experience: the Vision Pro's weight (600-650g) caused discomfort, while visual quality issues—darker rendering, color distortion (P3, P10, P11, P13, P19, P20, P23), insufficient resolution (P10, P26)—affected aesthetic appreciation. 

Despite these limitations, our controlled approach provided baseline evidence that perceptual modulation and conversational AI can enhance specific engagement dimensions: viewing duration, question diversity, and interpretive autonomy. Future research should progressively introduce authentic artworks, multiple simultaneous visitors, naturalistic conditions, and longer-term follow-up to assess ecological validity. These effects may be amplified by prior XR familiarity, which can reduce onboarding friction and increase tolerance to device constraints. Accordingly, the magnitude of change in engagement-related measures may differ for XR-unfamiliar visitors. Similarly, our evaluation prioritizes factual reliability over holistic interpretive quality, and should therefore be understood as a baseline assessment of system trustworthiness rather than a comprehensive curatorial evaluation.

\section{CONCLUSION}
This work has introduced WhiteTesseract, a high-resolution XR exhibition system designed to reshape how visitors engage with cultural heritage through adjustable sensory environments and conversational AI. Our study demonstrated that subtle perceptual modulation can significantly increase viewing time and foster focused attention, offering an alternative to spectacle-driven immersion in digital heritage applications. In parallel, our findings show that dialogic interaction—despite current limitations in LLMs—supports more autonomous and personally meaningful interpretations of visual heritage.

By integrating attentional design with interpretive flexibility, WhiteTesseract contributes to ongoing conversations in HCI and cultural heritage computing about how emerging technologies can augment, rather than replace, situated museum experiences. It presents a model for future systems that emphasize quiet adaptability, reflective engagement, and co-construction of meaning. Rather than pursuing complete automation, we argue for systems that sensitively mediate between technological possibilities and curatorial intentions, extending the interpretive potential of physical exhibitions while respecting their material and historical specificity.

\section*{Data Availability}

The experimental dialogue data generated in this study are publicly available on Zenodo at \url{https://doi.org/10.5281/zenodo.18335187}.
The repository includes all participant-generated questions, the corresponding system responses translated into English, and the original system responses in Chinese and Japanese, organized by artwork.
These materials allow direct inspection of the participant--LLM conversations and the qualitative factual ``marks'' used in the analysis, supporting transparency and reproducibility of the reported results.

\begin{acks}
This work was supported by JSPS KAKENHI Grant Number JP24K23878 (Grants-in-Aid for Scientific Research, 2024). We thank all participants of the user study.
\end{acks}

\bibliographystyle{ACM-Reference-Format}
\bibliography{sample-base}


\begin{thebibliography}{95}

%%% the Web address.

\ifx \showCODEN    \undefined \def \showCODEN     #1{\unskip}     \fi
\ifx \showDOI      \undefined \def \showDOI       #1{#1}\fi
\ifx \showISBNx    \undefined \def \showISBNx     #1{\unskip}     \fi
\ifx \showISBNxiii \undefined \def \showISBNxiii  #1{\unskip}     \fi
\ifx \showISSN     \undefined \def \showISSN      #1{\unskip}     \fi
\ifx \showLCCN     \undefined \def \showLCCN      #1{\unskip}     \fi
\ifx \shownote     \undefined \def \shownote      #1{#1}          \fi
\ifx \showarticletitle \undefined \def \showarticletitle #1{#1}   \fi
\ifx \showURL      \undefined \def \showURL       {\relax}        \fi
% invisible to TeX
\providecommand\bibfield[2]{#2}
\providecommand\bibinfo[2]{#2}
\providecommand\natexlab[1]{#1}
\providecommand\showeprint[2][]{arXiv:#2}

\bibitem[Achlioptas et~al\mbox{.}(2021)]%
        {achlioptas2021artemis}
\bibfield{author}{\bibinfo{person}{Panos Achlioptas}, \bibinfo{person}{Maks
  Ovsjanikov}, \bibinfo{person}{Kilichbek Haydarov}, \bibinfo{person}{Mohamed
  Elhoseiny}, {and} \bibinfo{person}{Leonidas~J Guibas}.}
  \bibinfo{year}{2021}\natexlab{}.
\newblock \showarticletitle{Artemis: Affective language for visual art}. In
  \bibinfo{booktitle}{\emph{Proceedings of the IEEE/CVF Conference on Computer
  Vision and Pattern Recognition}}. \bibinfo{pages}{11569--11579}.
\newblock


\bibitem[Ahmed et~al\mbox{.}(2025)]%
        {ahmed2025human}
\bibfield{author}{\bibinfo{person}{Nasim Ahmed}, \bibinfo{person}{Peng Wu},
  \bibinfo{person}{Kaiming Huang}, \bibinfo{person}{Sungchul Jung},
  \bibinfo{person}{Hansol Rheem}, \bibinfo{person}{Gang Tan},
  \bibinfo{person}{Mahdi Imani}, {and} \bibinfo{person}{Rifatul Islam}.}
  \bibinfo{year}{2025}\natexlab{}.
\newblock \showarticletitle{Human Task Performance and Associated Internal
  States in Extended Reality: A Systematic Review of Cognitive,
  Psychophysiological, and Physiological Dimensions}.
\newblock \bibinfo{journal}{\emph{Frontiers in Virtual Reality}}
  \bibinfo{volume}{6} (\bibinfo{year}{2025}), \bibinfo{pages}{1589256}.
\newblock


\bibitem[Alnagrat et~al\mbox{.}(2022)]%
        {alnagrat2022review}
\bibfield{author}{\bibinfo{person}{Ahmed Alnagrat},
  \bibinfo{person}{Rizalafande~Che Ismail}, \bibinfo{person}{Syed
  Zulkarnain~Syed Idrus}, {and} \bibinfo{person}{Rawad Mansour~Abdulhafith
  Alfaqi}.} \bibinfo{year}{2022}\natexlab{}.
\newblock \showarticletitle{A review of extended reality (XR) technologies in
  the future of human education: Current trend and future opportunity}.
\newblock \bibinfo{journal}{\emph{Journal of Human Centered Technology}}
  \bibinfo{volume}{1}, \bibinfo{number}{2} (\bibinfo{year}{2022}),
  \bibinfo{pages}{81--96}.
\newblock


\bibitem[Ausubel and Robinson(1969)]%
        {ausubel1969school}
\bibfield{author}{\bibinfo{person}{David~P Ausubel} {and}
  \bibinfo{person}{Floyd~G Robinson}.} \bibinfo{year}{1969}\natexlab{}.
\newblock \bibinfo{booktitle}{\emph{School learning: An introduction to
  educational psychology}}.
\newblock \bibinfo{publisher}{Holt, Rinehart Winston}.
\newblock


\bibitem[Bekele et~al\mbox{.}(2018)]%
        {bekele2018survey}
\bibfield{author}{\bibinfo{person}{Mafkereseb~Kassahun Bekele},
  \bibinfo{person}{Roberto Pierdicca}, \bibinfo{person}{Emanuele Frontoni},
  \bibinfo{person}{Eva~Savina Malinverni}, {and} \bibinfo{person}{James Gain}.}
  \bibinfo{year}{2018}\natexlab{}.
\newblock \showarticletitle{A survey of augmented, virtual, and mixed reality
  for cultural heritage}.
\newblock \bibinfo{journal}{\emph{Journal on Computing and Cultural Heritage
  (JOCCH)}} \bibinfo{volume}{11}, \bibinfo{number}{2} (\bibinfo{year}{2018}),
  \bibinfo{pages}{1--36}.
\newblock


\bibitem[Bender and Friedman(2018)]%
        {bender2018data}
\bibfield{author}{\bibinfo{person}{Emily~M Bender} {and} \bibinfo{person}{Batya
  Friedman}.} \bibinfo{year}{2018}\natexlab{}.
\newblock \showarticletitle{Data statements for natural language processing:
  Toward mitigating system bias and enabling better science}.
\newblock \bibinfo{journal}{\emph{Transactions of the Association for
  Computational Linguistics}}  \bibinfo{volume}{6} (\bibinfo{year}{2018}),
  \bibinfo{pages}{587--604}.
\newblock


\bibitem[Bender and Koller(2020)]%
        {bender2020climbing}
\bibfield{author}{\bibinfo{person}{Emily~M Bender} {and}
  \bibinfo{person}{Alexander Koller}.} \bibinfo{year}{2020}\natexlab{}.
\newblock \showarticletitle{Climbing towards NLU: On meaning, form, and
  understanding in the age of data}. In \bibinfo{booktitle}{\emph{Proceedings
  of the 58th annual meeting of the association for computational
  linguistics}}. \bibinfo{pages}{5185--5198}.
\newblock


\bibitem[Bennett(1995)]%
        {bennett1995birth}
\bibfield{author}{\bibinfo{person}{Tony Bennett}.}
  \bibinfo{year}{1995}\natexlab{}.
\newblock \showarticletitle{The Birth of the Museum: History, Theory,
  Politics}.
\newblock  (\bibinfo{year}{1995}).
\newblock


\bibitem[Bennett(2013)]%
        {bennett2013birth}
\bibfield{author}{\bibinfo{person}{Tony Bennett}.}
  \bibinfo{year}{2013}\natexlab{}.
\newblock \bibinfo{booktitle}{\emph{The birth of the museum: History, theory,
  politics}}.
\newblock \bibinfo{publisher}{Routledge}.
\newblock


\bibitem[Bhambra(2014)]%
        {bhambra2014postcolonial}
\bibfield{author}{\bibinfo{person}{Gurminder~K Bhambra}.}
  \bibinfo{year}{2014}\natexlab{}.
\newblock \showarticletitle{Postcolonial and decolonial dialogues}.
\newblock \bibinfo{journal}{\emph{Postcolonial studies}} \bibinfo{volume}{17},
  \bibinfo{number}{2} (\bibinfo{year}{2014}), \bibinfo{pages}{115--121}.
\newblock


\bibitem[Bharti(2024)]%
        {bhartiuseful}
\bibfield{author}{\bibinfo{person}{Mukesh~Chand Bharti}.}
  \bibinfo{year}{2024}\natexlab{}.
\newblock \showarticletitle{Useful online open-access digital resources and
  their selection for researchers in the digital information world}.
\newblock \bibinfo{journal}{\emph{IP Indian Journal of Library Science and
  Information Technology}} (\bibinfo{year}{2024}).
\newblock


\bibitem[Bhaskaran and Mahbub(2024)]%
        {bhaskaran2024immersive}
\bibfield{author}{\bibinfo{person}{Vasudev Bhaskaran} {and}
  \bibinfo{person}{Upal Mahbub}.} \bibinfo{year}{2024}\natexlab{}.
\newblock \showarticletitle{Immersive User Experiences: Trends and Challenges
  of Using XR Technologies}.
\newblock \bibinfo{journal}{\emph{Computer Vision}} (\bibinfo{year}{2024}),
  \bibinfo{pages}{260--278}.
\newblock


\bibitem[Billinghurst et~al\mbox{.}(2015)]%
        {billinghurst2015survey}
\bibfield{author}{\bibinfo{person}{Mark Billinghurst}, \bibinfo{person}{Adrian
  Clark}, \bibinfo{person}{Gun Lee}, {et~al\mbox{.}}}
  \bibinfo{year}{2015}\natexlab{}.
\newblock \showarticletitle{A survey of augmented reality}.
\newblock \bibinfo{journal}{\emph{Foundations and Trends{\textregistered} in
  Human--Computer Interaction}} \bibinfo{volume}{8}, \bibinfo{number}{2-3}
  (\bibinfo{year}{2015}), \bibinfo{pages}{73--272}.
\newblock


\bibitem[Biocca(1997)]%
        {biocca1997cyborg}
\bibfield{author}{\bibinfo{person}{Frank Biocca}.}
  \bibinfo{year}{1997}\natexlab{}.
\newblock \showarticletitle{The cyborg's dilemma: Progressive embodiment in
  virtual environments}.
\newblock \bibinfo{journal}{\emph{Journal of computer-mediated communication}}
  \bibinfo{volume}{3}, \bibinfo{number}{2} (\bibinfo{year}{1997}),
  \bibinfo{pages}{JCMC324}.
\newblock


\bibitem[Bishop(2005)]%
        {bishop2005installation}
\bibfield{author}{\bibinfo{person}{Claire Bishop}.}
  \bibinfo{year}{2005}\natexlab{}.
\newblock \showarticletitle{Installation art: A critical history}.
\newblock \bibinfo{journal}{\emph{(No Title)}} (\bibinfo{year}{2005}).
\newblock


\bibitem[Bitgood(2009)]%
        {bitgood2009museum}
\bibfield{author}{\bibinfo{person}{Stephen Bitgood}.}
  \bibinfo{year}{2009}\natexlab{}.
\newblock \showarticletitle{Museum fatigue: A critical review}.
\newblock \bibinfo{journal}{\emph{Visitor studies}} \bibinfo{volume}{12},
  \bibinfo{number}{2} (\bibinfo{year}{2009}), \bibinfo{pages}{93--111}.
\newblock


\bibitem[Bitgood(2016)]%
        {bitgood2016attention}
\bibfield{author}{\bibinfo{person}{Stephen Bitgood}.}
  \bibinfo{year}{2016}\natexlab{}.
\newblock \bibinfo{booktitle}{\emph{Attention and Value: Keys to understanding
  museum visitors}}.
\newblock \bibinfo{publisher}{Routledge}.
\newblock


\bibitem[Brieber et~al\mbox{.}(2015)]%
        {brieber2015white}
\bibfield{author}{\bibinfo{person}{David Brieber}, \bibinfo{person}{Marcos
  Nadal}, {and} \bibinfo{person}{Helmut Leder}.}
  \bibinfo{year}{2015}\natexlab{}.
\newblock \showarticletitle{In the white cube: Museum context enhances the
  valuation and memory of art}.
\newblock \bibinfo{journal}{\emph{Acta psychologica}}  \bibinfo{volume}{154}
  (\bibinfo{year}{2015}), \bibinfo{pages}{36--42}.
\newblock


\bibitem[Brieber et~al\mbox{.}(2014)]%
        {brieber2014art}
\bibfield{author}{\bibinfo{person}{David Brieber}, \bibinfo{person}{Marcos
  Nadal}, \bibinfo{person}{Helmut Leder}, {and} \bibinfo{person}{Raphael
  Rosenberg}.} \bibinfo{year}{2014}\natexlab{}.
\newblock \showarticletitle{Art in time and space: Context modulates the
  relation between art experience and viewing time}.
\newblock \bibinfo{journal}{\emph{PloS one}} \bibinfo{volume}{9},
  \bibinfo{number}{6} (\bibinfo{year}{2014}), \bibinfo{pages}{e99019}.
\newblock


\bibitem[Caramiaux(2023)]%
        {caramiaux2023ai}
\bibfield{author}{\bibinfo{person}{Baptiste Caramiaux}.}
  \bibinfo{year}{2023}\natexlab{}.
\newblock \showarticletitle{AI with Museums and Cultural Heritage}.
\newblock \bibinfo{journal}{\emph{AI in Museums: Reflections, Perspectives and
  Applications}} (\bibinfo{year}{2023}), \bibinfo{pages}{117--130}.
\newblock


\bibitem[Carbon(2017)]%
        {Carbon2017}
\bibfield{author}{\bibinfo{person}{Claus-Christian Carbon}.}
  \bibinfo{year}{2017}\natexlab{}.
\newblock \showarticletitle{Art Perception in the Museum: How We Spend Time and
  Space in Art Exhibitions}.
\newblock \bibinfo{journal}{\emph{i-Perception}} \bibinfo{volume}{8},
  \bibinfo{number}{1} (\bibinfo{year}{2017}).
\newblock
\urldef\tempurl%
\url{https://doi.org/10.1177/2041669517694184}
\showDOI{\tempurl}


\bibitem[Chen et~al\mbox{.}(2022)]%
        {chen2022pali}
\bibfield{author}{\bibinfo{person}{Xi Chen}, \bibinfo{person}{Xiao Wang},
  \bibinfo{person}{Soravit Changpinyo}, \bibinfo{person}{AJ Piergiovanni},
  \bibinfo{person}{Piotr Padlewski}, \bibinfo{person}{Daniel Salz},
  \bibinfo{person}{Sebastian Goodman}, \bibinfo{person}{Adam Grycner},
  \bibinfo{person}{Basil Mustafa}, \bibinfo{person}{Lucas Beyer},
  {et~al\mbox{.}}} \bibinfo{year}{2022}\natexlab{}.
\newblock \showarticletitle{Pali: A jointly-scaled multilingual language-image
  model}.
\newblock \bibinfo{journal}{\emph{arXiv preprint arXiv:2209.06794}}
  (\bibinfo{year}{2022}).
\newblock


\bibitem[Cheng et~al\mbox{.}(2022)]%
        {cheng2022towards}
\bibfield{author}{\bibinfo{person}{Yi~Fei Cheng}, \bibinfo{person}{Hang Yin},
  \bibinfo{person}{Yukang Yan}, \bibinfo{person}{Jan Gugenheimer}, {and}
  \bibinfo{person}{David Lindlbauer}.} \bibinfo{year}{2022}\natexlab{}.
\newblock \showarticletitle{Towards understanding diminished reality}. In
  \bibinfo{booktitle}{\emph{Proceedings of the 2022 CHI Conference on Human
  Factors in Computing Systems}}. \bibinfo{pages}{1--16}.
\newblock


\bibitem[Chi et~al\mbox{.}(2001)]%
        {chi2001learning}
\bibfield{author}{\bibinfo{person}{Michelene~TH Chi},
  \bibinfo{person}{Stephanie~A Siler}, \bibinfo{person}{Heisawn Jeong},
  \bibinfo{person}{Takashi Yamauchi}, {and} \bibinfo{person}{Robert~G
  Hausmann}.} \bibinfo{year}{2001}\natexlab{}.
\newblock \showarticletitle{Learning from human tutoring}.
\newblock \bibinfo{journal}{\emph{Cognitive science}} \bibinfo{volume}{25},
  \bibinfo{number}{4} (\bibinfo{year}{2001}), \bibinfo{pages}{471--533}.
\newblock


\bibitem[Christidou(2010)]%
        {christidou2010re}
\bibfield{author}{\bibinfo{person}{Dimitra Christidou}.}
  \bibinfo{year}{2010}\natexlab{}.
\newblock \showarticletitle{Re-Introducing Visitors: Thoughts and Discussion on
  John Falk’s Notion of Visitors’ Identity-Related Visit Motivations}.
\newblock \bibinfo{journal}{\emph{Papers from the Institute of Archaeology}}
  \bibinfo{volume}{20} (\bibinfo{year}{2010}), \bibinfo{pages}{111--122}.
\newblock


\bibitem[Chuah(2018)]%
        {chuah2018and}
\bibfield{author}{\bibinfo{person}{Stephanie Hui-Wen Chuah}.}
  \bibinfo{year}{2018}\natexlab{}.
\newblock \showarticletitle{Why and who will adopt extended reality technology?
  Literature review, synthesis, and future research agenda}.
\newblock \bibinfo{journal}{\emph{Literature Review, Synthesis, and Future
  Research Agenda (December 13, 2018)}} (\bibinfo{year}{2018}).
\newblock


\bibitem[Chun and Turk-Browne(2007)]%
        {chun2007interactions}
\bibfield{author}{\bibinfo{person}{Marvin~M Chun} {and}
  \bibinfo{person}{Nicholas~B Turk-Browne}.} \bibinfo{year}{2007}\natexlab{}.
\newblock \showarticletitle{Interactions between attention and memory}.
\newblock \bibinfo{journal}{\emph{Current opinion in neurobiology}}
  \bibinfo{volume}{17}, \bibinfo{number}{2} (\bibinfo{year}{2007}),
  \bibinfo{pages}{177--184}.
\newblock


\bibitem[{Collaboration for Ongoing Visitor Experience Studies (COVES)}(2025)]%
        {COVES2025}
\bibfield{author}{\bibinfo{person}{{Collaboration for Ongoing Visitor
  Experience Studies (COVES)}}.} \bibinfo{year}{2025}\natexlab{}.
\newblock \bibinfo{title}{Understanding Our Visitors: Multi-Institutional
  Museum Study (July 2024–June 2025)}.
\newblock
\newblock
\urldef\tempurl%
\url{https://understandingvisitors.org/images/downloads/Reports/coves_aggregate_report_2025_final_digital_single_pages.pdf}
\showURL{%
\tempurl}


\bibitem[Damala et~al\mbox{.}(2008)]%
        {damala2008bridging}
\bibfield{author}{\bibinfo{person}{Areti Damala}, \bibinfo{person}{Pierre
  Cubaud}, \bibinfo{person}{Anne Bationo}, \bibinfo{person}{Pascal Houlier},
  {and} \bibinfo{person}{Isabelle Marchal}.} \bibinfo{year}{2008}\natexlab{}.
\newblock \showarticletitle{Bridging the gap between the digital and the
  physical: design and evaluation of a mobile augmented reality guide for the
  museum visit}. In \bibinfo{booktitle}{\emph{Proceedings of the 3rd
  international conference on Digital Interactive Media in Entertainment and
  Arts}}. \bibinfo{pages}{120--127}.
\newblock


\bibitem[Darda et~al\mbox{.}(2025)]%
        {darda2025comparison}
\bibfield{author}{\bibinfo{person}{Kohinoor~M Darda}, \bibinfo{person}{Vicente
  Estrada~Gonzalez}, \bibinfo{person}{Alexander~P Christensen},
  \bibinfo{person}{Isabella Bobrow}, \bibinfo{person}{Amy Krimm},
  \bibinfo{person}{Zuha Nasim}, \bibinfo{person}{Eileen~R Cardillo},
  \bibinfo{person}{William Perthes}, {and} \bibinfo{person}{Anjan Chatterjee}.}
  \bibinfo{year}{2025}\natexlab{}.
\newblock \showarticletitle{A comparison of art engagement in museums and
  through digital media}.
\newblock \bibinfo{journal}{\emph{Scientific Reports}} \bibinfo{volume}{15},
  \bibinfo{number}{1} (\bibinfo{year}{2025}), \bibinfo{pages}{8972}.
\newblock


\bibitem[Dierking and Falk(1992)]%
        {dierking1992redefining}
\bibfield{author}{\bibinfo{person}{Lynn~D Dierking} {and}
  \bibinfo{person}{John~H Falk}.} \bibinfo{year}{1992}\natexlab{}.
\newblock \showarticletitle{Redefining the museum experience: the interactive
  experience model}.
\newblock \bibinfo{journal}{\emph{Visitor Studies}} \bibinfo{volume}{4},
  \bibinfo{number}{1} (\bibinfo{year}{1992}), \bibinfo{pages}{173--176}.
\newblock


\bibitem[Dourish(2001)]%
        {dourish2001action}
\bibfield{author}{\bibinfo{person}{Paul Dourish}.}
  \bibinfo{year}{2001}\natexlab{}.
\newblock \bibinfo{booktitle}{\emph{Where the action is: the foundations of
  embodied interaction}}.
\newblock \bibinfo{publisher}{MIT press}.
\newblock


\bibitem[Falk(2016)]%
        {Falk2016Motivations}
\bibfield{author}{\bibinfo{person}{John~H. Falk}.}
  \bibinfo{year}{2016}\natexlab{}.
\newblock \bibinfo{booktitle}{\emph{Understanding Museum Visitors’
  Motivations and Learning}}.
\newblock \bibinfo{type}{{T}echnical {R}eport}. \bibinfo{institution}{Danish
  Agency for Culture}.
\newblock
\urldef\tempurl%
\url{https://slks.dk/fileadmin/user_upload/dokumenter/KS/institutioner/museer/Indsatsomraader/Brugerundersoegelse/Artikler/John_Falk_Understanding_museum_visitors__motivations_and_learning.pdf}
\showURL{%
\tempurl}


\bibitem[Falk and Dierking(2016)]%
        {falk2016museum}
\bibfield{author}{\bibinfo{person}{John~H Falk} {and} \bibinfo{person}{Lynn~D
  Dierking}.} \bibinfo{year}{2016}\natexlab{}.
\newblock \bibinfo{booktitle}{\emph{The museum experience revisited}}.
\newblock \bibinfo{publisher}{Routledge}.
\newblock


\bibitem[Falk and Dierking(2018)]%
        {falk2018learning}
\bibfield{author}{\bibinfo{person}{John~H Falk} {and} \bibinfo{person}{Lynn~D
  Dierking}.} \bibinfo{year}{2018}\natexlab{}.
\newblock \bibinfo{booktitle}{\emph{Learning from museums}}.
\newblock \bibinfo{publisher}{Rowman \& Littlefield}.
\newblock


\bibitem[Forrest(2013)]%
        {forrest2013museum}
\bibfield{author}{\bibinfo{person}{Regan Forrest}.}
  \bibinfo{year}{2013}\natexlab{}.
\newblock \showarticletitle{Museum atmospherics: The role of the exhibition
  environment in the visitor experience}.
\newblock \bibinfo{journal}{\emph{Visitor Studies}} \bibinfo{volume}{16},
  \bibinfo{number}{2} (\bibinfo{year}{2013}), \bibinfo{pages}{201--216}.
\newblock


\bibitem[Gabriel(2023)]%
        {gabriel2023socially}
\bibfield{author}{\bibinfo{person}{Saadia Gabriel}.}
  \bibinfo{year}{2023}\natexlab{}.
\newblock \bibinfo{booktitle}{\emph{Socially Responsible and Factual Reasoning
  for Equitable AI Systems}}.
\newblock \bibinfo{publisher}{University of Washington}.
\newblock


\bibitem[Gaia et~al\mbox{.}(2019)]%
        {gaia2019engaging}
\bibfield{author}{\bibinfo{person}{Giuliano Gaia}, \bibinfo{person}{Stefania
  Boiano}, {and} \bibinfo{person}{Ann Borda}.} \bibinfo{year}{2019}\natexlab{}.
\newblock \showarticletitle{Engaging museum visitors with AI: The case of
  chatbots}.
\newblock In \bibinfo{booktitle}{\emph{Museums and Digital Culture: New
  Perspectives and Research}}. \bibinfo{publisher}{Springer},
  \bibinfo{pages}{309--329}.
\newblock


\bibitem[Garbutt et~al\mbox{.}(2020)]%
        {garbutt2020embodied}
\bibfield{author}{\bibinfo{person}{Michael Garbutt}, \bibinfo{person}{Scott
  East}, \bibinfo{person}{Branka Spehar}, \bibinfo{person}{Vicente
  Estrada-Gonzalez}, \bibinfo{person}{Brooke Carson-Ewart}, {and}
  \bibinfo{person}{Josephine Touma}.} \bibinfo{year}{2020}\natexlab{}.
\newblock \showarticletitle{The embodied gaze: Exploring applications for
  mobile eye tracking in the art museum}.
\newblock \bibinfo{journal}{\emph{Visitor Studies}} \bibinfo{volume}{23},
  \bibinfo{number}{1} (\bibinfo{year}{2020}), \bibinfo{pages}{82--100}.
\newblock


\bibitem[Gibbs~Jr(2005)]%
        {gibbs2005embodiment}
\bibfield{author}{\bibinfo{person}{Raymond~W Gibbs~Jr}.}
  \bibinfo{year}{2005}\natexlab{}.
\newblock \bibinfo{booktitle}{\emph{Embodiment and cognitive science}}.
\newblock \bibinfo{publisher}{Cambridge University Press}.
\newblock


\bibitem[Greenberg(2005)]%
        {greenberg2005vital}
\bibfield{author}{\bibinfo{person}{Stephen Greenberg}.}
  \bibinfo{year}{2005}\natexlab{}.
\newblock \showarticletitle{The vital museum}.
\newblock In \bibinfo{booktitle}{\emph{Reshaping museum space}}.
  \bibinfo{publisher}{Routledge}, \bibinfo{pages}{226--237}.
\newblock


\bibitem[Groys(2008)]%
        {groys2008art}
\bibfield{author}{\bibinfo{person}{Boris Groys}.}
  \bibinfo{year}{2008}\natexlab{}.
\newblock \bibinfo{booktitle}{\emph{Art power}}. Vol.~\bibinfo{volume}{8}.
\newblock \bibinfo{publisher}{MIT press Cambridge, MA}.
\newblock


\bibitem[Gunatilake et~al\mbox{.}(2024)]%
        {gunatilake2024enablers}
\bibfield{author}{\bibinfo{person}{Hashini Gunatilake}, \bibinfo{person}{John
  Grundy}, \bibinfo{person}{Rashina Hoda}, {and} \bibinfo{person}{Ingo
  Mueller}.} \bibinfo{year}{2024}\natexlab{}.
\newblock \showarticletitle{Enablers and barriers of empathy in software
  developer and user interactions: a mixed methods case study}.
\newblock \bibinfo{journal}{\emph{ACM Transactions on Software Engineering and
  Methodology}} \bibinfo{volume}{33}, \bibinfo{number}{4}
  (\bibinfo{year}{2024}), \bibinfo{pages}{1--41}.
\newblock


\bibitem[Hashemi and Kamps(2018)]%
        {hashemi2018exploiting}
\bibfield{author}{\bibinfo{person}{Seyyed~Hadi Hashemi} {and}
  \bibinfo{person}{Jaap Kamps}.} \bibinfo{year}{2018}\natexlab{}.
\newblock \showarticletitle{Exploiting behavioral user models for point of
  interest recommendation in smart museums}.
\newblock \bibinfo{journal}{\emph{New Review of Hypermedia and Multimedia}}
  \bibinfo{volume}{24}, \bibinfo{number}{3} (\bibinfo{year}{2018}),
  \bibinfo{pages}{228--261}.
\newblock


\bibitem[Ho et~al\mbox{.}(2025)]%
        {ho2025enhancing}
\bibfield{author}{\bibinfo{person}{Hoang~Phuoc Ho}, \bibinfo{person}{Vani
  Ramesh}, \bibinfo{person}{Ivo Zaloudek}, \bibinfo{person}{Delaram~Javdani
  Rikhtehgar}, {and} \bibinfo{person}{Shenghui Wang}.}
  \bibinfo{year}{2025}\natexlab{}.
\newblock \showarticletitle{Enhancing Visitor Engagement in Interactive Art
  Exhibitions with Visual-Enhanced Conversational Agents}. In
  \bibinfo{booktitle}{\emph{Proceedings of the 30th International Conference on
  Intelligent User Interfaces}}. \bibinfo{pages}{660--671}.
\newblock


\bibitem[Hooper-Greenhill(2020)]%
        {hooper2020museums}
\bibfield{author}{\bibinfo{person}{Eilean Hooper-Greenhill}.}
  \bibinfo{year}{2020}\natexlab{}.
\newblock \bibinfo{booktitle}{\emph{Museums and the interpretation of visual
  culture}}.
\newblock \bibinfo{publisher}{Routledge}.
\newblock


\bibitem[Kahneman(1973)]%
        {kahneman1973attention}
\bibfield{author}{\bibinfo{person}{Daniel Kahneman}.}
  \bibinfo{year}{1973}\natexlab{}.
\newblock \bibinfo{booktitle}{\emph{Attention and effort}}.
  Vol.~\bibinfo{volume}{1063}.
\newblock \bibinfo{publisher}{Citeseer}.
\newblock


\bibitem[Kailas and Tiwari(2021)]%
        {kailas2021design}
\bibfield{author}{\bibinfo{person}{Ganesh Kailas} {and}
  \bibinfo{person}{Nachiketa Tiwari}.} \bibinfo{year}{2021}\natexlab{}.
\newblock \showarticletitle{Design for immersive experience: Role of spatial
  audio in extended reality applications}. In \bibinfo{booktitle}{\emph{Design
  for Tomorrow—Volume 2: Proceedings of ICoRD 2021}}. Springer,
  \bibinfo{pages}{853--863}.
\newblock


\bibitem[Kenthapadi et~al\mbox{.}(2024)]%
        {kenthapadi2024grounding}
\bibfield{author}{\bibinfo{person}{Krishnaram Kenthapadi},
  \bibinfo{person}{Mehrnoosh Sameki}, {and} \bibinfo{person}{Ankur Taly}.}
  \bibinfo{year}{2024}\natexlab{}.
\newblock \showarticletitle{Grounding and evaluation for large language models:
  Practical challenges and lessons learned (survey)}. In
  \bibinfo{booktitle}{\emph{Proceedings of the 30th ACM SIGKDD Conference on
  Knowledge Discovery and Data Mining}}. \bibinfo{pages}{6523--6533}.
\newblock


\bibitem[Kim and Baylor(2006)]%
        {kim2006social}
\bibfield{author}{\bibinfo{person}{Yanghee Kim} {and} \bibinfo{person}{Amy~L
  Baylor}.} \bibinfo{year}{2006}\natexlab{}.
\newblock \showarticletitle{A social-cognitive framework for pedagogical agents
  as learning companions}.
\newblock \bibinfo{journal}{\emph{Educational technology research and
  development}}  \bibinfo{volume}{54} (\bibinfo{year}{2006}),
  \bibinfo{pages}{569--596}.
\newblock


\bibitem[Kim et~al\mbox{.}(2006)]%
        {kim2006pedagogical}
\bibfield{author}{\bibinfo{person}{Yanghee Kim}, \bibinfo{person}{Amy~L
  Baylor}, {and} \bibinfo{person}{Pals Group}.}
  \bibinfo{year}{2006}\natexlab{}.
\newblock \showarticletitle{Pedagogical agents as learning companions: The role
  of agent competency and type of interaction}.
\newblock \bibinfo{journal}{\emph{Educational technology research and
  development}}  \bibinfo{volume}{54} (\bibinfo{year}{2006}),
  \bibinfo{pages}{223--243}.
\newblock


\bibitem[Laugwitz et~al\mbox{.}(2008)]%
        {laugwitz2008construction}
\bibfield{author}{\bibinfo{person}{Bettina Laugwitz}, \bibinfo{person}{Theo
  Held}, {and} \bibinfo{person}{Martin Schrepp}.}
  \bibinfo{year}{2008}\natexlab{}.
\newblock \showarticletitle{Construction and evaluation of a user experience
  questionnaire}. In \bibinfo{booktitle}{\emph{Symposium of the Austrian HCI
  and usability engineering group}}. Springer, \bibinfo{pages}{63--76}.
\newblock


\bibitem[Lee and Kim(2025)]%
        {lee2025diminishar}
\bibfield{author}{\bibinfo{person}{JangHyeon Lee} {and}
  \bibinfo{person}{Lawrence~H Kim}.} \bibinfo{year}{2025}\natexlab{}.
\newblock \showarticletitle{DiminishAR: Diminishing Visual Distractions via
  Holographic AR Displays}. In \bibinfo{booktitle}{\emph{Proceedings of the
  2025 CHI Conference on Human Factors in Computing Systems}}.
  \bibinfo{pages}{1--16}.
\newblock


\bibitem[Leinhardt et~al\mbox{.}(2003)]%
        {leinhardt2003learning}
\bibfield{author}{\bibinfo{person}{Gaea Leinhardt}, \bibinfo{person}{Kevin
  Crowley}, {and} \bibinfo{person}{Karen Knutson}.}
  \bibinfo{year}{2003}\natexlab{}.
\newblock \bibinfo{booktitle}{\emph{Learning conversations in museums}}.
\newblock \bibinfo{publisher}{Taylor \& Francis}.
\newblock


\bibitem[Leinhardt and Knutson(2004)]%
        {leinhardt2004listening}
\bibfield{author}{\bibinfo{person}{Gaea Leinhardt} {and} \bibinfo{person}{Karen
  Knutson}.} \bibinfo{year}{2004}\natexlab{}.
\newblock \bibinfo{booktitle}{\emph{Listening in on museum conversations}}.
\newblock \bibinfo{publisher}{Rowman Altamira}.
\newblock


\bibitem[Liang et~al\mbox{.}(2024)]%
        {liang2024effect}
\bibfield{author}{\bibinfo{person}{Hsin-Yi Liang}, \bibinfo{person}{Gwo-Jen
  Hwang}, \bibinfo{person}{Tien-Yu Hsu}, {and} \bibinfo{person}{Jen-Yuan Yeh}.}
  \bibinfo{year}{2024}\natexlab{}.
\newblock \showarticletitle{Effect of an AI-based chatbot on students' learning
  performance in alternate reality game-based museum learning}.
\newblock \bibinfo{journal}{\emph{British Journal of Educational Technology}}
  \bibinfo{volume}{55}, \bibinfo{number}{5} (\bibinfo{year}{2024}),
  \bibinfo{pages}{2315--2338}.
\newblock


\bibitem[Lindblom et~al\mbox{.}(2025)]%
        {lindblom2025take}
\bibfield{author}{\bibinfo{person}{Jessica Lindblom}, \bibinfo{person}{Julia
  Ros{\'e}n}, \bibinfo{person}{Maurice Lamb}, {and} \bibinfo{person}{Erik
  Billing}.} \bibinfo{year}{2025}\natexlab{}.
\newblock \showarticletitle{“Take Nothing on its Look”: Revealing Users’
  Expectations and Experiences in Social Human-Robot Interaction}.
\newblock \bibinfo{journal}{\emph{ACM Transactions on Human-Robot Interaction}}
  (\bibinfo{year}{2025}).
\newblock


\bibitem[Lonetree(2012)]%
        {lonetree2012decolonizing}
\bibfield{author}{\bibinfo{person}{Amy Lonetree}.}
  \bibinfo{year}{2012}\natexlab{}.
\newblock \bibinfo{booktitle}{\emph{Decolonizing museums: Representing Native
  America in national and tribal museums}}.
\newblock \bibinfo{publisher}{Univ of North Carolina Press}.
\newblock


\bibitem[Luger and Sellen(2016)]%
        {luger2016like}
\bibfield{author}{\bibinfo{person}{Ewa Luger} {and} \bibinfo{person}{Abigail
  Sellen}.} \bibinfo{year}{2016}\natexlab{}.
\newblock \showarticletitle{" Like Having a Really Bad PA" The Gulf between
  User Expectation and Experience of Conversational Agents}. In
  \bibinfo{booktitle}{\emph{Proceedings of the 2016 CHI conference on human
  factors in computing systems}}. \bibinfo{pages}{5286--5297}.
\newblock


\bibitem[Machidon et~al\mbox{.}(2020)]%
        {machidon2020culturalerica}
\bibfield{author}{\bibinfo{person}{Octavian-Mihai Machidon},
  \bibinfo{person}{Ale{\v{s}} Tav{\v{c}}ar}, \bibinfo{person}{Matja{\v{z}}
  Gams}, {and} \bibinfo{person}{Mihai Dugulean{\u{a}}}.}
  \bibinfo{year}{2020}\natexlab{}.
\newblock \showarticletitle{CulturalERICA: A conversational agent improving the
  exploration of European cultural heritage}.
\newblock \bibinfo{journal}{\emph{Journal of Cultural Heritage}}
  \bibinfo{volume}{41} (\bibinfo{year}{2020}), \bibinfo{pages}{152--165}.
\newblock


\bibitem[Makransky and Petersen(2021)]%
        {makransky2021cognitive}
\bibfield{author}{\bibinfo{person}{Guido Makransky} {and}
  \bibinfo{person}{Gustav~B Petersen}.} \bibinfo{year}{2021}\natexlab{}.
\newblock \showarticletitle{The cognitive affective model of immersive learning
  (CAMIL): A theoretical research-based model of learning in immersive virtual
  reality}.
\newblock \bibinfo{journal}{\emph{Educational psychology review}}
  \bibinfo{volume}{33}, \bibinfo{number}{3} (\bibinfo{year}{2021}),
  \bibinfo{pages}{937--958}.
\newblock


\bibitem[Maxwell and Evans(2002)]%
        {maxwell2002museums}
\bibfield{author}{\bibinfo{person}{Lorraine~E Maxwell} {and}
  \bibinfo{person}{Gary~W Evans}.} \bibinfo{year}{2002}\natexlab{}.
\newblock \showarticletitle{Museums as learning settings: The importance of the
  physical environment}.
\newblock \bibinfo{journal}{\emph{Journal of Museum Education}}
  \bibinfo{volume}{27}, \bibinfo{number}{1} (\bibinfo{year}{2002}),
  \bibinfo{pages}{3--7}.
\newblock


\bibitem[McTear(2022)]%
        {mctear2022conversational}
\bibfield{author}{\bibinfo{person}{Michael McTear}.}
  \bibinfo{year}{2022}\natexlab{}.
\newblock \bibinfo{booktitle}{\emph{Conversational ai: Dialogue systems,
  conversational agents, and chatbots}}.
\newblock \bibinfo{publisher}{Springer Nature}.
\newblock


\bibitem[Mori et~al\mbox{.}(2017)]%
        {mori2017survey}
\bibfield{author}{\bibinfo{person}{Shohei Mori}, \bibinfo{person}{Sei Ikeda},
  {and} \bibinfo{person}{Hideo Saito}.} \bibinfo{year}{2017}\natexlab{}.
\newblock \showarticletitle{A survey of diminished reality: Techniques for
  visually concealing, eliminating, and seeing through real objects}.
\newblock \bibinfo{journal}{\emph{IPSJ Transactions on Computer Vision and
  Applications}}  \bibinfo{volume}{9} (\bibinfo{year}{2017}),
  \bibinfo{pages}{1--14}.
\newblock


\bibitem[Naveed et~al\mbox{.}(2023)]%
        {naveed2023comprehensive}
\bibfield{author}{\bibinfo{person}{Humza Naveed}, \bibinfo{person}{Asad~Ullah
  Khan}, \bibinfo{person}{Shi Qiu}, \bibinfo{person}{Muhammad Saqib},
  \bibinfo{person}{Saeed Anwar}, \bibinfo{person}{Muhammad Usman},
  \bibinfo{person}{Naveed Akhtar}, \bibinfo{person}{Nick Barnes}, {and}
  \bibinfo{person}{Ajmal Mian}.} \bibinfo{year}{2023}\natexlab{}.
\newblock \showarticletitle{A comprehensive overview of large language models}.
\newblock \bibinfo{journal}{\emph{arXiv preprint arXiv:2307.06435}}
  (\bibinfo{year}{2023}).
\newblock


\bibitem[No{\"e}(2004)]%
        {noe2004action}
\bibfield{author}{\bibinfo{person}{Alva No{\"e}}.}
  \bibinfo{year}{2004}\natexlab{}.
\newblock \bibinfo{booktitle}{\emph{Action in perception}}.
\newblock \bibinfo{publisher}{MIT press}.
\newblock


\bibitem[O'brien et~al\mbox{.}(2012)]%
        {o2012understanding}
\bibfield{author}{\bibinfo{person}{Marita~A O'brien}, \bibinfo{person}{Wendy~A
  Rogers}, {and} \bibinfo{person}{Arthur~D Fisk}.}
  \bibinfo{year}{2012}\natexlab{}.
\newblock \showarticletitle{Understanding age and technology experience
  differences in use of prior knowledge for everyday technology interactions}.
\newblock \bibinfo{journal}{\emph{ACM Transactions on Accessible Computing
  (TACCESS)}} \bibinfo{volume}{4}, \bibinfo{number}{2} (\bibinfo{year}{2012}),
  \bibinfo{pages}{1--27}.
\newblock


\bibitem[O'doherty(1999)]%
        {o1999inside}
\bibfield{author}{\bibinfo{person}{Brian O'doherty}.}
  \bibinfo{year}{1999}\natexlab{}.
\newblock \bibinfo{booktitle}{\emph{Inside the white cube: The ideology of the
  gallery space}}.
\newblock \bibinfo{publisher}{Univ of California Press}.
\newblock


\bibitem[O'Doherty(2007)]%
        {o2007studio}
\bibfield{author}{\bibinfo{person}{Brian O'Doherty}.}
  \bibinfo{year}{2007}\natexlab{}.
\newblock \bibinfo{booktitle}{\emph{Studio and cube: on the relationship
  between where art is made and where art is displayed}}.
  Vol.~\bibinfo{volume}{1}.
\newblock \bibinfo{publisher}{Princeton Architectural Press}.
\newblock


\bibitem[Posner(1980)]%
        {posner1980orienting}
\bibfield{author}{\bibinfo{person}{Michael~I Posner}.}
  \bibinfo{year}{1980}\natexlab{}.
\newblock \showarticletitle{Orienting of attention}.
\newblock \bibinfo{journal}{\emph{Quarterly journal of experimental
  psychology}} \bibinfo{volume}{32}, \bibinfo{number}{1}
  (\bibinfo{year}{1980}), \bibinfo{pages}{3--25}.
\newblock


\bibitem[Radianti et~al\mbox{.}(2020)]%
        {radianti2020systematic}
\bibfield{author}{\bibinfo{person}{Jaziar Radianti}, \bibinfo{person}{Tim~A
  Majchrzak}, \bibinfo{person}{Jennifer Fromm}, {and} \bibinfo{person}{Isabell
  Wohlgenannt}.} \bibinfo{year}{2020}\natexlab{}.
\newblock \showarticletitle{A systematic review of immersive virtual reality
  applications for higher education: Design elements, lessons learned, and
  research agenda}.
\newblock \bibinfo{journal}{\emph{Computers \& education}}
  \bibinfo{volume}{147} (\bibinfo{year}{2020}), \bibinfo{pages}{103778}.
\newblock


\bibitem[Rogoff(2013)]%
        {rogoff2013terra}
\bibfield{author}{\bibinfo{person}{Irit Rogoff}.}
  \bibinfo{year}{2013}\natexlab{}.
\newblock \bibinfo{booktitle}{\emph{Terra infirma: Geography's visual
  culture}}.
\newblock \bibinfo{publisher}{Routledge}.
\newblock


\bibitem[Rosemary(2025)]%
        {rosemary2025leveraging}
\bibfield{author}{\bibinfo{person}{Falola~Titilope Rosemary}.}
  \bibinfo{year}{2025}\natexlab{}.
\newblock \showarticletitle{Leveraging Artificial Intelligence and Data
  Analytics for Enhancing Museum Experiences: Exploring Historical Narratives,
  Visitor Engagement, and Digital Transformation in the Age of Innovation}.
\newblock \bibinfo{journal}{\emph{Int Res J Mod Eng Technol Sci}}
  \bibinfo{volume}{7}, \bibinfo{number}{1} (\bibinfo{year}{2025}).
\newblock


\bibitem[Ruan et~al\mbox{.}(2019)]%
        {ruan2019quizbot}
\bibfield{author}{\bibinfo{person}{Sherry Ruan}, \bibinfo{person}{Liwei Jiang},
  \bibinfo{person}{Justin Xu}, \bibinfo{person}{Bryce Joe-Kun Tham},
  \bibinfo{person}{Zhengneng Qiu}, \bibinfo{person}{Yeshuang Zhu},
  \bibinfo{person}{Elizabeth~L Murnane}, \bibinfo{person}{Emma Brunskill},
  {and} \bibinfo{person}{James~A Landay}.} \bibinfo{year}{2019}\natexlab{}.
\newblock \showarticletitle{Quizbot: A dialogue-based adaptive learning system
  for factual knowledge}. In \bibinfo{booktitle}{\emph{Proceedings of the 2019
  CHI conference on human factors in computing systems}}.
  \bibinfo{pages}{1--13}.
\newblock


\bibitem[Schaffer et~al\mbox{.}(2021)]%
        {schaffer2021questions}
\bibfield{author}{\bibinfo{person}{Stefan Schaffer}, \bibinfo{person}{Aaron
  Ru{\ss}}, \bibinfo{person}{Mino~Lee Sasse}, \bibinfo{person}{Louise
  Schubotz}, {and} \bibinfo{person}{Oliver Gustke}.}
  \bibinfo{year}{2021}\natexlab{}.
\newblock \showarticletitle{Questions and answers: Important steps to let AI
  chatbots answer questions in the museum}. In
  \bibinfo{booktitle}{\emph{International Conference on ArtsIT, Interactivity
  and Game Creation}}. Springer, \bibinfo{pages}{346--358}.
\newblock


\bibitem[Serafin et~al\mbox{.}(2020)]%
        {serafin2020reflections}
\bibfield{author}{\bibinfo{person}{Stefania Serafin}, \bibinfo{person}{Federico
  Avanzini}, \bibinfo{person}{Amalia De~Goetzen}, \bibinfo{person}{Cumhur
  Erkut}, \bibinfo{person}{Michele Geronazzo}, \bibinfo{person}{Francesco
  Grani}, \bibinfo{person}{Niels~Christian Nilsson}, {and}
  \bibinfo{person}{Rolf Nordahl}.} \bibinfo{year}{2020}\natexlab{}.
\newblock \showarticletitle{Reflections from five years of Sonic Interactions
  in Virtual Environments workshops}.
\newblock \bibinfo{journal}{\emph{Journal of New Music Research}}
  \bibinfo{volume}{49}, \bibinfo{number}{1} (\bibinfo{year}{2020}),
  \bibinfo{pages}{24--34}.
\newblock


\bibitem[Serrell(2015)]%
        {serrell2015exhibit}
\bibfield{author}{\bibinfo{person}{Beverly Serrell}.}
  \bibinfo{year}{2015}\natexlab{}.
\newblock \bibinfo{booktitle}{\emph{Exhibit labels: An interpretive approach}}.
\newblock \bibinfo{publisher}{Rowman \& Littlefield}.
\newblock


\bibitem[Shuster et~al\mbox{.}(2022)]%
        {shuster2022blenderbot}
\bibfield{author}{\bibinfo{person}{Kurt Shuster}, \bibinfo{person}{Jing Xu},
  \bibinfo{person}{Mojtaba Komeili}, \bibinfo{person}{Da Ju},
  \bibinfo{person}{Eric~Michael Smith}, \bibinfo{person}{Stephen Roller},
  \bibinfo{person}{Megan Ung}, \bibinfo{person}{Moya Chen},
  \bibinfo{person}{Kushal Arora}, \bibinfo{person}{Joshua Lane},
  {et~al\mbox{.}}} \bibinfo{year}{2022}\natexlab{}.
\newblock \showarticletitle{Blenderbot 3: a deployed conversational agent that
  continually learns to responsibly engage}.
\newblock \bibinfo{journal}{\emph{arXiv preprint arXiv:2208.03188}}
  (\bibinfo{year}{2022}).
\newblock


\bibitem[Slater and Wilbur(1997)]%
        {slater1997framework}
\bibfield{author}{\bibinfo{person}{Mel Slater} {and} \bibinfo{person}{Sylvia
  Wilbur}.} \bibinfo{year}{1997}\natexlab{}.
\newblock \showarticletitle{A framework for immersive virtual environments
  (FIVE): Speculations on the role of presence in virtual environments}.
\newblock \bibinfo{journal}{\emph{Presence: Teleoperators \& Virtual
  Environments}} \bibinfo{volume}{6}, \bibinfo{number}{6}
  (\bibinfo{year}{1997}), \bibinfo{pages}{603--616}.
\newblock


\bibitem[Smith and Smith(2001)]%
        {Smith2001}
\bibfield{author}{\bibinfo{person}{James~K. Smith} {and}
  \bibinfo{person}{Lisa~F. Smith}.} \bibinfo{year}{2001}\natexlab{}.
\newblock \showarticletitle{Spending Time on Art}.
\newblock \bibinfo{journal}{\emph{Empirical Studies of the Arts}}
  \bibinfo{volume}{19}, \bibinfo{number}{2} (\bibinfo{year}{2001}),
  \bibinfo{pages}{229--236}.
\newblock
\urldef\tempurl%
\url{https://doi.org/10.2190/5MQM-59JH-X21R-JN5J}
\showDOI{\tempurl}


\bibitem[Styliani et~al\mbox{.}(2009)]%
        {styliani2009virtual}
\bibfield{author}{\bibinfo{person}{Sylaiou Styliani},
  \bibinfo{person}{Liarokapis Fotis}, \bibinfo{person}{Kotsakis Kostas}, {and}
  \bibinfo{person}{Patias Petros}.} \bibinfo{year}{2009}\natexlab{}.
\newblock \showarticletitle{Virtual museums, a survey and some issues for
  consideration}.
\newblock \bibinfo{journal}{\emph{Journal of cultural Heritage}}
  \bibinfo{volume}{10}, \bibinfo{number}{4} (\bibinfo{year}{2009}),
  \bibinfo{pages}{520--528}.
\newblock


\bibitem[Team(2024)]%
        {gemini_flash}
\bibfield{author}{\bibinfo{person}{Gemini Team}.}
  \bibinfo{year}{2024}\natexlab{}.
\newblock \bibinfo{title}{Gemini 1.5: Unlocking multimodal understanding across
  millions of tokens of context}.
\newblock
\newblock
\showeprint[arxiv]{2403.05530}~[cs.CL]
\urldef\tempurl%
\url{https://arxiv.org/abs/2403.05530}
\showURL{%
\tempurl}


\bibitem[tom Dieck and Jung(2017)]%
        {tom2017value}
\bibfield{author}{\bibinfo{person}{M~Claudia tom Dieck} {and}
  \bibinfo{person}{Timothy~Hyungsoo Jung}.} \bibinfo{year}{2017}\natexlab{}.
\newblock \showarticletitle{Value of augmented reality at cultural heritage
  sites: A stakeholder approach}.
\newblock \bibinfo{journal}{\emph{Journal of destination marketing \&
  management}} \bibinfo{volume}{6}, \bibinfo{number}{2} (\bibinfo{year}{2017}),
  \bibinfo{pages}{110--117}.
\newblock


\bibitem[Treisman and Gelade(1980)]%
        {treisman1980feature}
\bibfield{author}{\bibinfo{person}{Anne~M Treisman} {and}
  \bibinfo{person}{Garry Gelade}.} \bibinfo{year}{1980}\natexlab{}.
\newblock \showarticletitle{A feature-integration theory of attention}.
\newblock \bibinfo{journal}{\emph{Cognitive psychology}} \bibinfo{volume}{12},
  \bibinfo{number}{1} (\bibinfo{year}{1980}), \bibinfo{pages}{97--136}.
\newblock


\bibitem[Tsitseklis et~al\mbox{.}(2023)]%
        {tsitseklis2023recbot}
\bibfield{author}{\bibinfo{person}{Konstantinos Tsitseklis},
  \bibinfo{person}{Georgia Stavropoulou}, \bibinfo{person}{Anastasios
  Zafeiropoulos}, \bibinfo{person}{Athina Thanou}, {and}
  \bibinfo{person}{Symeon Papavassiliou}.} \bibinfo{year}{2023}\natexlab{}.
\newblock \showarticletitle{RECBOT: Virtual Museum navigation through a Chatbot
  assistant and personalized Recommendations}. In
  \bibinfo{booktitle}{\emph{Adjunct proceedings of the 31st ACM conference on
  user modeling, adaptation and personalization}}. \bibinfo{pages}{388--396}.
\newblock


\bibitem[Tzortzi et~al\mbox{.}(2017)]%
        {tzortzi2017rethinking}
\bibfield{author}{\bibinfo{person}{Kali Tzortzi} {et~al\mbox{.}}}
  \bibinfo{year}{2017}\natexlab{}.
\newblock \showarticletitle{Rethinking museum space: interaction between
  spatial layout design and digital sensory environments}. Space Syntax
  Symposium.
\newblock


\bibitem[VanLehn(2011)]%
        {vanlehn2011relative}
\bibfield{author}{\bibinfo{person}{Kurt VanLehn}.}
  \bibinfo{year}{2011}\natexlab{}.
\newblock \showarticletitle{The relative effectiveness of human tutoring,
  intelligent tutoring systems, and other tutoring systems}.
\newblock \bibinfo{journal}{\emph{Educational psychologist}}
  \bibinfo{volume}{46}, \bibinfo{number}{4} (\bibinfo{year}{2011}),
  \bibinfo{pages}{197--221}.
\newblock


\bibitem[Varitimiadis et~al\mbox{.}(2021)]%
        {varitimiadis2021graph}
\bibfield{author}{\bibinfo{person}{Savvas Varitimiadis},
  \bibinfo{person}{Konstantinos Kotis}, \bibinfo{person}{Dimitra Pittou}, {and}
  \bibinfo{person}{Georgios Konstantakis}.} \bibinfo{year}{2021}\natexlab{}.
\newblock \showarticletitle{Graph-based conversational AI: Towards a
  distributed and collaborative multi-chatbot approach for museums}.
\newblock \bibinfo{journal}{\emph{Applied Sciences}} \bibinfo{volume}{11},
  \bibinfo{number}{19} (\bibinfo{year}{2021}), \bibinfo{pages}{9160}.
\newblock


\bibitem[Varitimiadis et~al\mbox{.}(2020)]%
        {varitimiadis2020towards}
\bibfield{author}{\bibinfo{person}{Savvas Varitimiadis},
  \bibinfo{person}{Konstantinos Kotis}, \bibinfo{person}{Andreas Skamagis},
  \bibinfo{person}{Alexandros Tzortzakakis}, \bibinfo{person}{George
  Tsekouras}, {and} \bibinfo{person}{Dimitris Spiliotopoulos}.}
  \bibinfo{year}{2020}\natexlab{}.
\newblock \showarticletitle{Towards implementing an AI chatbot platform for
  museums}. In \bibinfo{booktitle}{\emph{International conference on cultural
  informatics, communication \& media studies}}, Vol.~\bibinfo{volume}{1}.
\newblock


\bibitem[Vom~Lehn and Heath(2005)]%
        {vom2005accounting}
\bibfield{author}{\bibinfo{person}{Dirk Vom~Lehn} {and}
  \bibinfo{person}{Christian Heath}.} \bibinfo{year}{2005}\natexlab{}.
\newblock \showarticletitle{Accounting for new technology in museum
  exhibitions}.
\newblock \bibinfo{journal}{\emph{International Journal of Arts Management}}
  (\bibinfo{year}{2005}), \bibinfo{pages}{11--21}.
\newblock


\bibitem[Vom~Lehn et~al\mbox{.}(2001)]%
        {vom2001exhibiting}
\bibfield{author}{\bibinfo{person}{Dirk Vom~Lehn}, \bibinfo{person}{Christian
  Heath}, {and} \bibinfo{person}{Jon Hindmarsh}.}
  \bibinfo{year}{2001}\natexlab{}.
\newblock \showarticletitle{Exhibiting interaction: Conduct and collaboration
  in museums and galleries}.
\newblock \bibinfo{journal}{\emph{Symbolic interaction}} \bibinfo{volume}{24},
  \bibinfo{number}{2} (\bibinfo{year}{2001}), \bibinfo{pages}{189--216}.
\newblock


\bibitem[Weech et~al\mbox{.}(2019)]%
        {weech2019presence}
\bibfield{author}{\bibinfo{person}{S{\'e}amas Weech}, \bibinfo{person}{Sophie
  Kenny}, {and} \bibinfo{person}{Michael Barnett-Cowan}.}
  \bibinfo{year}{2019}\natexlab{}.
\newblock \showarticletitle{Presence and cybersickness in virtual reality are
  negatively related: a review}.
\newblock \bibinfo{journal}{\emph{Frontiers in psychology}}
  \bibinfo{volume}{10} (\bibinfo{year}{2019}), \bibinfo{pages}{158}.
\newblock


\bibitem[White et~al\mbox{.}(2023)]%
        {promptPattern}
\bibfield{author}{\bibinfo{person}{Jules White}, \bibinfo{person}{Quchen Fu},
  \bibinfo{person}{Sam Hays}, \bibinfo{person}{Michael Sandborn},
  \bibinfo{person}{Carlos Olea}, \bibinfo{person}{Henry Gilbert},
  \bibinfo{person}{Ashraf Elnashar}, \bibinfo{person}{Jesse Spencer-Smith},
  {and} \bibinfo{person}{Douglas~C. Schmidt}.} \bibinfo{year}{2023}\natexlab{}.
\newblock \showarticletitle{A Prompt Pattern Catalog to Enhance Prompt
  Engineering with ChatGPT}. In \bibinfo{booktitle}{\emph{Proceedings of the
  30th Conference on Pattern Languages of Programs}} (Monticello, IL, USA)
  \emph{(\bibinfo{series}{PLoP '23})}. \bibinfo{publisher}{The Hillside Group},
  \bibinfo{address}{USA}, Article \bibinfo{articleno}{5},
  \bibinfo{numpages}{31}~pages.
\newblock
\showISBNx{9781941652190}


\bibitem[Wickens et~al\mbox{.}(2022)]%
        {wickens2022applied}
\bibfield{author}{\bibinfo{person}{Christopher~D Wickens},
  \bibinfo{person}{Jason~S McCarley}, {and} \bibinfo{person}{Robert~S
  Gutzwiller}.} \bibinfo{year}{2022}\natexlab{}.
\newblock \bibinfo{booktitle}{\emph{Applied attention theory}}.
\newblock \bibinfo{publisher}{CRC press}.
\newblock


\bibitem[Wood et~al\mbox{.}(2006)]%
        {wood2006attention}
\bibfield{author}{\bibinfo{person}{Sharon Wood}, \bibinfo{person}{Richard Cox},
  {and} \bibinfo{person}{Peter Cheng}.} \bibinfo{year}{2006}\natexlab{}.
\newblock \showarticletitle{Attention design: Eight issues to consider}.
\newblock \bibinfo{journal}{\emph{Computers in Human Behavior}}
  \bibinfo{volume}{22}, \bibinfo{number}{4} (\bibinfo{year}{2006}),
  \bibinfo{pages}{588--602}.
\newblock


\end{thebibliography}

\appendix

\section{Participant Demographics}

To contextualize the results of our study, we collected demographic information and background variables from all 26 participants prior to the experimental session. This included age group, gender, educational background, domain of study, prior exposure to art exhibitions, and familiarity with AR/VR/MR technologies. The questionnaire also asked about participants' expectations regarding the use of AR in art experiences. Table~\ref{tab:participant_demographics} summarizes key participant characteristics and confirms a technically experienced, XR-familiar sample, which we interpret as an early-adopter population rather than a representative museum audience.

\begin{table}[H]
\centering
\caption{Demographic Overview of Participants (N=26)}
\begin{tabularx}{\textwidth}{>{\raggedright\arraybackslash\textbf}l X r r}
\toprule
\textbf{Attribute} & \textbf{Category} & \textbf{N} & \textbf{\%} \\
\midrule
\textbf{Age group} & 20–25 & 12 & 46 \\
          & 25–30 & 14 & 54 \\
\textbf{Gender} & Male & 18 & 70 \\
          & Female & 8 & 30 \\
\textbf{Educational Level} & Undergraduate & 3 & 12 \\
                  & Master’s & 10 & 38 \\
                  & Doctoral & 10 & 38 \\
                  & Other (e.g., researchers, pre-graduate) & 3 & 12 \\
\textbf{Academic Background} & Engineering \& Technology & 24 & 92 \\
                    & Business \& Management & 1 & 4 \\
                    & Not specified & 1 & 4 \\
\textbf{Frequency of Art Exhibition Visit} & Rarely (1–2 times/year) & 11 & 42 \\
                                  & Occasionally (every few months) & 12 & 46 \\
                                  & Frequently (monthly or more) & 3 & 12 \\
\textbf{Knowledge of Art History} & None / Minimal (1–2) & 11 & 42 \\
                         & Moderate (3) & 12 & 46 \\
                         & High (4–5) & 3 & 12 \\
\textbf{Preferred Art Forms*} & Digital/Media Art & 24 & 92 \\
                    & Painting & 17 & 65 \\
                    & Installation & 15 & 58 \\
                    & Photography & 11 & 42 \\
                    & Sculpture & 6 & 23 \\
\textbf{Experience with AR/VR/MR} & Yes & 26 & 100 \\
\textbf{Frequency of AR/VR/MR Use} & Rarely (1–2 times/year) & 14 & 54 \\
                       & Occasionally (monthly) & 10 & 38 \\
                       & Frequently (weekly) & 2 & 8 \\
\textbf{Main Purpose of Use*} & Entertainment (movie/games) & 22 & 85 \\
                    & Education / Research & 14 & 54 \\
                    & Work / Development & 5 & 19 \\
\textbf{Expectations for AR-enhanced Art Experience*} & Learning Enhancement & 24 & 92 \\
                                             & Immersive Engagement & 16 & 62 \\
                                             & Transformative Experience & 13 & 50 \\
                                             & Technical Innovation & 15 & 58 \\
                                             & Minimal Enhancement & 2 & 8 \\
\bottomrule
\end{tabularx}
\label{tab:participant_demographics}
\begin{tablenotes}
\small
\item \textit{Note: All questionnaire items were mandatory and included both single-choice and multiple-choice questions. Items marked with an asterisk (\textasteriskcentered) indicate questions allowing multiple responses.}
\end{tablenotes}
\end{table}

\section{Post-Experiment Questionnaires}

To evaluate participants’ experiences and perceptions after interacting with the WhiteTesseract system, we administered a set of post-experiment questionnaires that included both standardized and custom-designed items. First, we adapted the User Experience Questionnaire (UEQ) to assess participants’ general impressions of the system across pragmatic and hedonic dimensions such as clarity, supportiveness, and inventiveness (Table~\ref{tab:ueq}). Next, a customized set of items was designed to assess perceptions of the XR system's impact on aesthetic appreciation, immersion, comfort, and attentional modulation in the art exhibition context (Table~\ref{tab:post_experiment_questionnaire}). Finally, a separate questionnaire focused on participants’ engagement with the LLM(AI-powered voice guide), specifically examining perceived learning value, adaptability, and interaction style (Table~\ref{tab:llm_questionnaire}). All items were rated using a 5-point Likert scale.

\begin{table}[ht]
\centering
\caption{Adapted User Experience Questionnaire (UEQ) for WhiteTesseract system.}
\begin{tabular}{lll}
\toprule
\textbf{No.} & \textbf{Scale} & \textbf{Adjective Pair} \\
\midrule
Q1 & 1--5 & Obstructive --- Supportive \\
Q2 & 1--5 & Complicated --- Easy \\
Q3 & 1--5 & Inefficient --- Efficient \\
Q4 & 1--5 & Confusing --- Clear \\
Q5 & 1--5 & Boring --- Exciting \\
Q6 & 1--5 & Not Interesting --- Interesting \\
Q7 & 1--5 & Conventional --- Inventive \\
Q8 & 1--5 & Usual --- Leading Edge \\
\bottomrule
\end{tabular}
\label{tab:ueq}
\begin{tablenotes}
\small
\item \textit{Note: Participants were asked to evaluate their experience with the WhiteTesseract system using the above bipolar adjective scales. Each item was rated on a 5-point Likert scale, where 1 represents the left adjective, and 5 represents the right.}
\end{tablenotes}
\end{table}

\begin{table}[ht]
\centering
\caption{XR Technology and Art Experience Evaluation Questionnaire Items}
\begin{tabularx}{\textwidth}{@{}p{0.05\textwidth} p{0.9\textwidth}@{}}
\toprule
\textbf{No.} & \textbf{Question Text} \\
\midrule
Q1 & I am very satisfied with the overall experience provided by the AR’s visual
focusing effect. \\
Q2 & The visual quality of the AR overlays enhanced my appreciation of the
artworks. \\
Q3 & The AR content was accurately placed relative to the physical artworks,
improving my understanding. \\
Q4 & The AR system's responsiveness to my movements felt natural and
unobtrusive. \\
Q5 & Wearing the AR device throughout the exhibition was comfortable and did not
distract from the art viewing experience. \\
Q6 & The visual isolation feature effectively helped me focus on individual
artworks. \\
Q7 & The noise reduction feature significantly improved my immersion in the
exhibition. \\
Q8 & The AR technology made me spend more time with each artwork than I
would have in a traditional exhibition. \\
Q9 & I gained deeper insights into the artworks through the AR features than I
would have without them. \\
Q10 & Overall, the AR technology enhanced my art viewing experience, and I would
prefer future exhibitions to incorporate similar AR features. \\
\bottomrule
\end{tabularx}
\label{tab:post_experiment_questionnaire}
\begin{tablenotes}
\small
\item \textit{Note: All items were rated on a 5-point Likert scale (1 = Strongly Disagree, 5 = Strongly Agree).}
\end{tablenotes}
\end{table}

\begin{table}[ht]
\centering
\caption{The LLM (AI-powered voice guide) Engagement and Learning Style Questionnaire Items}
\begin{tabularx}{\textwidth}{@{}p{0.05\textwidth} p{0.9\textwidth}@{}}
\toprule
\textbf{No.} & \textbf{Question Text} \\
\midrule
Q1 & I am very satisfied with the overall experience provided by the AI voice
interaction. \\
Q2 & The AI-powered voice guide provided valuable insights that deepened my
understanding of the artworks. \\
Q3 & Because of the LLM dialog system, I spent more time on each artwork. \\
Q4 & Did the LLM system inspire you to ask questions you might not have thought of otherwise? \\
Q5 & Did the LLM system adapt well to your level of art knowledge? \\
Q6 & Would you prefer to have this LLM dialog system available in future art
exhibitions? \\
\bottomrule
\end{tabularx}
\label{tab:llm_questionnaire}
\begin{tablenotes}
\small
\item \textit{Note: All items were rated on a 5-point Likert scale (1 = Strongly Disagree, 5 = Strongly Agree).}
\end{tablenotes}
\end{table}

\section{Semi-structured Interview Questions}

To complement the quantitative findings, we conducted semi-structured interviews with all participants after their interaction with the WhiteTesseract system. These interviews aimed to gain deeper insights into participants’ subjective impressions, perceived impact of the XR features, and their preferences for future exhibition formats. Participants were encouraged to reflect on their experiences, challenges, and design suggestions in their own words, allowing for open-ended responses and the emergence of nuanced themes. The full set of interview prompts used in this study is listed in Table~\ref{tab:Semi-structured}.

\begin{table}[ht]
\centering
\caption{Semi-structured Interviews' Questionnaire Items}
\begin{tabularx}{\textwidth}{@{}p{0.05\textwidth} p{0.9\textwidth}@{}}
\toprule
\textbf{No.} & \textbf{Question Text} \\
\midrule
Q1 & Can you describe your overall experience with the WhiteTesseract system? \\
Q2 & How do you feel the XR technology changed your interaction with the artworks compared to a traditional exhibition? \\
Q3 & Were there any moments during the experience that particularly stood out to you, either positively or negatively? (Were there any features of the XR system that you found particularly challenging or distracting?) \\
Q4 & How did the physical presence of the XR device affect your movement and comfort in the exhibition space? \\
Q5 & If you could design your ideal WhiteTesseract system, what features or experiences would you include? \\
Q6 & If you could choose your favorite type of exhibition among a 'fully digital experience,' a 'fully physical experience,' and a 'combination of digital and physical experiences,' which one would you pick? Which do you think is more suitable for future development? \\
\bottomrule
\end{tabularx}
\label{tab:Semi-structured}
\end{table}

\end{document}